\documentclass[twocolumn,english,pra,aps,superscriptaddress,floatfix]{revtex4-1}

\usepackage{amsthm}
\usepackage{amsmath}
\usepackage{graphicx}
\usepackage{amssymb}
\usepackage{bm}
\usepackage{latexsym}
\graphicspath{{paperimages/}}

\makeatletter
\@ifundefined{textcolor}{}
{%
 \definecolor{BLACK}{gray}{0}
 \definecolor{WHITE}{gray}{1}
 \definecolor{RED}{rgb}{1,0,0}
 \definecolor{GREEN}{rgb}{0,1,0}
 \definecolor{BLUE}{rgb}{0,0,1}
 \definecolor{CYAN}{cmyk}{1,0,0,0}
 \definecolor{MAGENTA}{cmyk}{0,1,0,0}
 \definecolor{YELLOW}{cmyk}{0,0,1,0}
 }

\newcommand{\di}{\partial}

\makeatother

\usepackage{babel}

\begin{document}

\author{F. Hasan}
\affiliation{Department of Physics and Astronomy, McMaster University, 1280 Main St.\ W., Hamilton, ON, L8S 4M1, Canada}
\author{D. H. J. O'Dell}
\affiliation{Department of Physics and Astronomy, McMaster University, 1280 Main St.\ W., Hamilton, ON, L8S 4M1, Canada}

\date{\today}

\title{Parametric amplification of light in a cavity with a moving dielectric membrane: Landau-Zener problem for the Maxwell field }

\begin{abstract}
We perform a theoretical investigation into the classical and quantum dynamics of an optical field in a cavity containing a moving membrane (``membrane-in-the-middle'' set-up). Our approach is based on the Maxwell wave equation, and complements previous studies based on an effective Hamiltonian. The analysis shows that for slowly moving and weakly reflective membranes the dynamics can be approximated by unitary, first-order-in-time evolution given by an effective Schr\"{o}dinger-like equation with a Hamiltonian that does not depend on the membrane speed. This approximate theory is the one typically adopted in cavity optomechanics and we develop a criterion for its validity. However, in more general situations the full second-order wave equation predicts light dynamics which do not conserve energy, giving rise to parametric amplification (or reduction) that is forbidden under first order dynamics and can be considered to be the classical counterpart of the dynamical Casimir effect. The case of a membrane moving at constant velocity can be mapped onto the Landau-Zener problem but with additional terms responsible for field amplification. Furthermore, the nature of the adiabatic regime is rather different from the ordinary Schr\"{o}dinger  case, since mode amplitudes need not be constant even when there are no transitions between them. The Landau-Zener problem for a field is therefore richer than in the standard single-particle case.  We use the work-energy theorem applied to the radiation pressure on the membrane as a self-consistency check for our solutions of the wave equation and as a tool to gain an intuitive understanding of energy pumped into/out of the light field by the motion of the membrane.  

\end{abstract}


\maketitle

\section{Introduction}
\label{sec:intro}

Most textbooks on quantum optics (see, e.g.\ \cite{Heitler,Loudon,Walls+Milburn,Mandel+Wolf}) begin with Maxwell's equations and use them to obtain a wave equation for the field which is second order in time and space. The normal modes of this equation behave like independent harmonic oscillators and can be quantized by the methods of ordinary non-relativistic quantum mechanics.
In this way, the quantum dynamics of the electromagnetic field is shown to be governed by the Schr\"{o}dinger equation which is first order in time and hence unitary (the lack of Lorentz invariance in Schr\"{o}dinger's equation should not worry us because normal modes separate time and space \cite{Tong}). This standard procedure breaks down in the presence of moving mirrors or dielectrics because there are no normal modes in time dependent systems.

Our mission in this paper is to study the nature of the dynamics, especially adiabaticity and parametric amplification, for an optical field in the presence of a moving dielectric in a cavity. In the absence of true normal modes we use time-evolving modes which become coupled, an approach inspired by the papers of C. K. Law \cite{Law1,Law2,Law3}. 
We are primarily interested in classical fields, however, we are naturally led to a comparison with the quantum case because  under certain approximations the time-evolving classical modes obey first-order equations which are mathematically analogous to Schr\"{o}dinger equations. The differences between first and second order wave equations have been previously studied in the context of the Klein-Gordon equation where it is known that the wave function cannot be interpreted as a probability amplitude, in contrast to that of the Schr\"{o}dinger equation \cite{Greiner}. Indeed, the Klein-Gordon equation does not provide a consistent description of a single particle precisely because it allows particle creation and annihilation (the Klein-Gordon equation does, however, correctly describe the normal modes of a free spinless quantum field). Similarly, in the dynamical Casimir effect (DCE) pairs of photons are generated from the vacuum by a moving mirror \cite{Dodonov09,Dalvit10}, and here we study the classical analogue of this phenomenon in the form of parametric amplification.

A well known form of the DCE is Davies-Fulling-DeWitt radiation \cite{Davies75,Fulling76,Dewitt}  generated in response to the uniform acceleration of a single mirror in free space. It is related to the Unruh effect  \cite{Unruh}, and therefore ultimately to Hawking radiation \cite{Hawking}.  The DCE in a cavity with a moving end mirror was first investigated by Moore in 1970 \cite{Moore}. If the mirror is oscillated at twice the frequency of a cavity mode the condition for parametric resonance is fulfilled and the effect is exponentially enhanced \cite{Dodonov95,Dalvit99,Plunien,Schaller}. Still, the effect is tiny and various schemes have been devised to enhance or mimic it.  When a gas or semiconductor is ionized to produce a plasma the refractive index can drop to near zero in a picosecond  \cite{Yablonovitch,Manko}, and when the ionization is produced by a periodically pulsed laser the result can be a rapidly oscillating plasma mirror \cite{Lozovik,Crocce,Braggio05}. Similarly, a coherently pumped $\chi^{(2)}$ nonlinear crystal forms an optical parametric oscillator whose nonlinear susceptibility oscillates at optical frequencies  \cite{Lambrecht10}.  The first system to successfully observe the DCE operated in the microwave regime and used a superconducting circuit made of a coplanar transmission line, the effective length of which can be changed at frequencies exceeding 10 GHz by modulating the inductance \cite{Wilson,Johansson}.

The interaction of light with a moving dielectric is a rich problem whose history goes back at least as far as the investigations carried out by Fresnel \cite{Fresnel} and Fizeau \cite{Fizeau} in the 19th century. It has close connections to the theory of special relativity, and, in the case of nonuniform motion, to general relativity \cite{PiwnickiLeonhardt}. An active modern area of research that involves moving dielectrics is the field of optomechanics \cite{Kippenberg+Vahala07,AspelmeyerKippenbergMarquardt}, where light and mechanical oscillators are coupled through radiation pressure. The prototypical system consists of a cavity made of two mirrors, one of which is mounted on a spring. When pumped by a laser, the optical field that builds up inside the cavity can displace the mobile mirror by radiation pressure. Such a set-up was realized in 1983 by Dorsel \textit{et al} \cite{dorsel83}  who observed a lengthening of the cavity. The dynamic version of this effect, where the mirror position and light field amplitude oscillate, can be used to heat or cool the mirror motion, as first demonstrated by Braginsky and co-workers in experiments with microwave cavities \cite{braginsky67,braginsky70} in the 1960s.   The past decade has seen renewed theoretical \cite{KippenbergRae,GirvinMarquardt,AspelmeyerGenes,FreegardeXuereb} and experimental \cite{AspelmeyerZeilinger,arcizet06,Schliesser06,VahalaRokhsari,KarraiFavero,Schliesser08,thompson08,HarrisJayich,HarrisZwickl,Painter1,Groblacher09,Park09,Schliesser09,Rocheleau10,Sankey2010,KippenbergRiviere,Chan11,LehnertRegal1,HarrisLee} activity in optomechanics, with one of the principal aims being to laser cool a mechanical object towards its quantum ground state. In particular, the experiment \cite{Chan11} achieved a sub-single phonon occupancy of a nanomechanical oscillator.  Optomechanical systems have now been realized in diverse physical media including ultrahigh-Q microtoroids \cite{VahalaRokhsari}, mirrors attached to cantilevers \cite{KarraiFavero,AspelmeyerZeilinger}, optomechanical crystals \cite{Painter1}, mechanical oscillators in microwave and optical cavities \cite{LehnertRegal1},  cold atom clouds \cite{murch08,brennecke08}, hybrid atom-membrane optomechanics \cite{Treutlein1,MeystreBariani}, as well as the  `membrane-in-the-middle' cavities \cite{thompson08,HarrisJayich,HarrisZwickl,Sankey2010,HarrisLee} that will be the focus of this paper. Radiation pressure and its quantum fluctuations (shot noise) on mirrors also turn out to be significant issues in high precision optical interferometers, like those designed to detect gravitational waves \cite{Braginsky+Manukin,braginsky01,braginsky02,corbitt06}.

\begin{figure}
 \includegraphics[width=0.9\columnwidth]{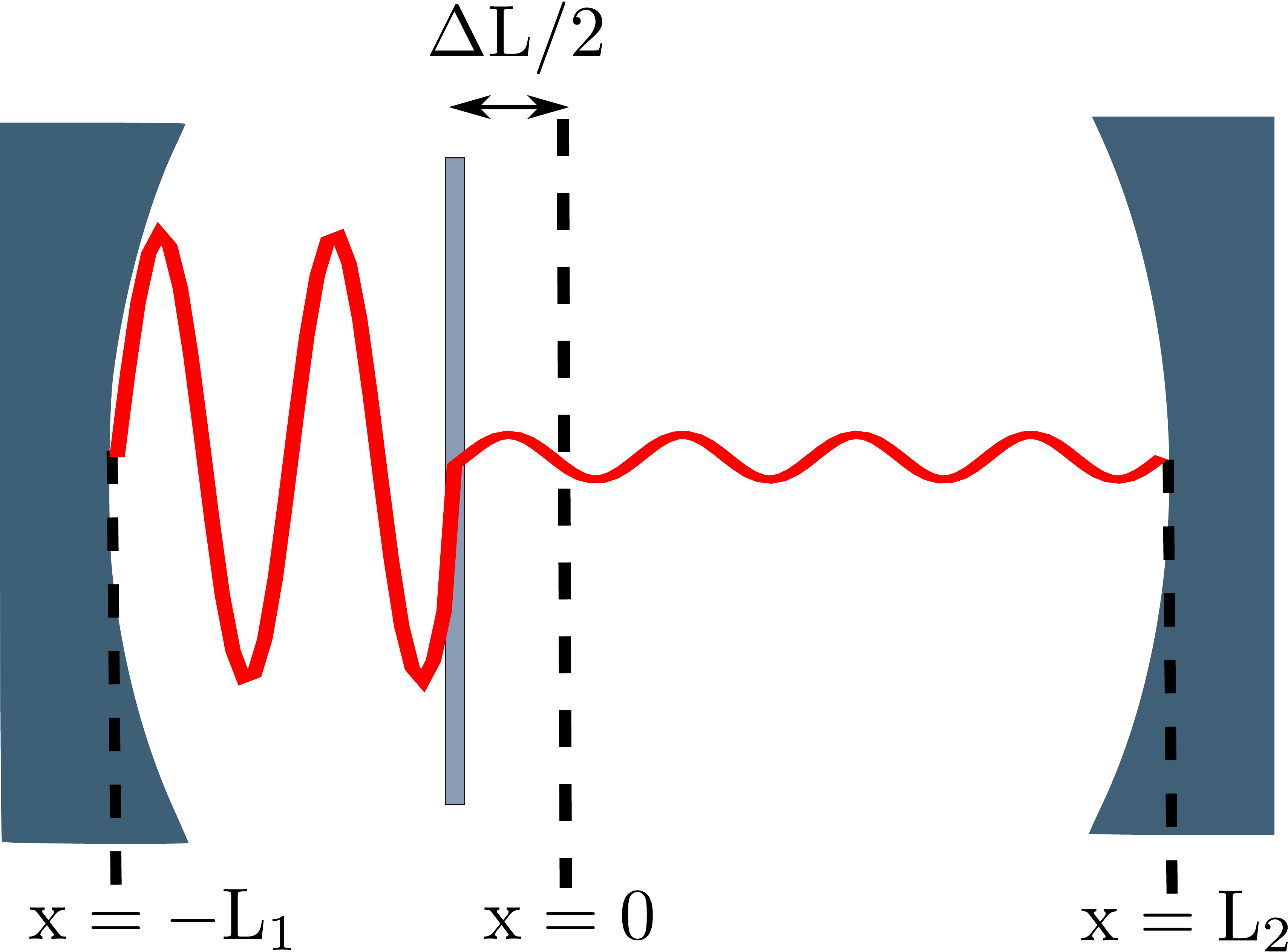}
 \caption{Schematic of the amplitude of light in a double cavity with perfectly reflective end mirrors and a partially transmissive, moveable central membrane.}
 \label{fig:DoubleCavityPic}
 \end{figure}

In this paper we investigate the dynamics of light stored in a  `membrane-in-the-middle' type optical cavity, as depicted schematically in Figure \ref{fig:DoubleCavityPic}. This arrangement has been realized in a series of experiments by the Yale group \cite{thompson08,HarrisJayich,HarrisZwickl,Sankey2010,HarrisLee}, and is made of two highly-reflective end mirrors between which a thin moveable membrane (slab of SiN dielectric approximately 50 nm thick) is suspended, forming two subcavities. Light is transmitted between the two cavities at a rate determined by the membrane reflectivity: when its reflectivity is high the membrane strongly alters the optical mode structure of the cavity producing a network of avoided crossings as a function of membrane displacement (see Figure \ref{fig:AvoidedCrossing}). The quadratic form of the mode structure at an avoided crossing lends itself to a quantum non-demolition measurement of the membrane's energy and hence a fundamental demonstration of the quantization of the energy of a mechanical oscillator, something which is not possible with linear coupling \cite{Sankey2010,HarrisLee}. Nonclassical correlations between two mechanical modes in such membranes has also been demonstrated experimentally \cite{Patil15,Vengalatorre15}.

  The small gaps between the optical modes at avoided crossings in a membrane-in-the-middle cavity mean that such systems have a fundamentally multi-mode character. This has led other authors \cite{HarrisPhotonShuttle}, as well as us \cite{NickPaper}, to suggest that Landau-Zener type physics might be relevant to the optical dynamics caused by membrane motion. The celebrated  Landau-Zener problem is one of the few exactly solvable problems in time-dependent quantum mechanics and provides a paradigm for analyzing the dynamical control of quantum systems, including the breakdown of adiabatic transfer between states. Applying this to the electromagnetic field where photon number is not conserved is one of the main themes of this paper.  In our previous paper \cite{NickPaper} we showed how to approximately map the dynamics of two interacting \emph{classical} optical fields obeying the Maxwell wave equation in the membrane-in-the-middle cavity system onto the \emph{mathematics} of the Landau-Zener model, and hence how to analyze the efficiency of light transfer from one subcavity to the other by moving the membrane. Such deterministic transfer of light between cavities is a basic element of a quantum network \cite{Kimble}, and has technological significance for cavity-QED realizations of quantum information processing \cite{Roadmap}.

The optomechanical interaction between a mirror and a cavity mode of frequency $\omega_{\mathrm{cav}}$ arises from radiation pressure and is usually written \cite{AspelmeyerKippenbergMarquardt,HarrisJayich,HarrisPhotonShuttle,MarquardtHeinrich2}
\begin{eqnarray}
\hat{H}_{\mathrm{optomech}} & = & \hbar \omega_{\mathrm{cav}}(x)  \hat{a}^{\dag} \hat{a} \\
& = &  \hbar   \left(\omega_{\mathrm{cav}} + x \frac{ \partial \omega_{\mathrm{cav}}}{\partial x} + x^2 \frac{ \partial^2 \omega_{\mathrm{cav}}}{\partial x^2} \ldots \right) \hat{a}^{\dag} \hat{a} \nonumber
\label{eq:Hoptomech}
\end{eqnarray}
where $x$ is the mirror displacement from equilibrium and $\hat{a}^{\dag} \hat{a}$ gives the number of photons in the cavity mode. This interaction depends parametrically on mirror displacement through the dependence of the mode frequency $\omega_{\mathrm{cav}}$ on $x$. For small displacements in comparison to the optical wavelength it is sufficient to expand $\omega_{\mathrm{cav}}$ as shown; for the case of a single cavity with a mobile end mirror only the linear term is needed, but in a double cavity near an avoided crossing this vanishes and the leading term is quadratic.  In currently experimentally accessible regimes this Hamiltonian gives an excellent description. Nevertheless, $H_{\mathrm{optomech}}$ as written above has \emph{no dependence on the mirror's speed} (for either the exact or expanded form). This means that the DCE is excluded which is unsatisfactory form a purely theoretical standpoint. A more complete Hamiltonian for the double cavity which does include the DCE has been derived by Law \cite{Law3} and will be discussed in Section \ref{sec:DCE}. However, rather than using a Hamiltonian, our approach here will be based on the Maxwell wave equation. A similar approach to the one we take has recently been used by Casta\~{n}os and Weder \cite{WederCastanos} to describe a single cavity with a mobile end mirror; they combine Maxwell's wave equation for the light with Newton's equation for the mobile mirror. We, on the other hand, give the membrane a prescribed trajectory in order to make full contact with the Landau-Zener problem.

The DCE is a phenomenon that originates in quantum zero-point fluctuations of the vacuum  \cite{Jaekel},  and so does not occur in the classical field description. Nevertheless, there are classical analogues to the creation/annihilation of photons in the form of parametric amplification/reduction of pre-existing classical fields by time-dependent cavity boundaries  \cite{Askaryan}, however even this is absent in the standard Landau-Zener model applied previously to the membrane-in-the-middle system in references \cite{HarrisPhotonShuttle,NickPaper} because it obeys unitary time evolution. The approximate mapping from the Maxwell wave equation to an effective Schr\"{o}dinger-like wave equation has the unintended consequence of treating  classical field amplitudes as though they were probability amplitudes; whereas the sum of the squares of classical field amplitudes is proportional to the total energy and is not constrained to be constant in a time-dependent cavity, the sum of the squares of probability amplitudes is fixed at unity even for time-dependent Hamiltonians. The second-order-in-time nature of the Maxwell wave equation thus allows for a richer dynamical behaviour than is present in the standard Landau-Zener problem. We shall see that the second-order-in-time dynamics includes a type of evolution which is adiabatic in the sense that there is no transfer  (scattering) of light between modes and yet the magnitudes can still change due to parametric amplification/reduction, something which cannot occur in unitary evolution. In this paper we shall specifically investigate how such ``beyond Landau-Zener''  phenomena depend on membrane speed and reflectivity.

Nonunitary effects such as parametric amplification of the cavity field are negligible in standard optomechanical experiments but interesting from a fundamental perspective. In order to evaluate the prospects of observing them  it is important to know what membrane speeds can be achieved. One way to move the membrane in a prescribed motion such as a Landau-Zener sweep is to use a piezoelectric motor (as is used, for example, to stabilize cavities against vibrations \cite{Ottl}). The maximum speed would then be around 10 m/s. However, much greater effective speeds can be achieved without moving the membrane at all, but by instead filling the subcavities with dielectrics whose indices of refraction can be changed independently in time, thereby changing their relative optical lengths, see Appendix C of \cite{NickPaper}. Ultrafast electro-optical control of the refractive index allows the effective optical length of a cavity to be changed on time scales shorter than 10 ps \cite{LipsonPreble,LipsonDong,Reed,Yanik,LipsonXu07}. This control can be achieved by using a laser to excite a plasma of free charge in the dielectric, similar to the original proposal in reference \cite{Yablonovitch} mentioned above. Related effects can also be generated electrically \cite{LipsonXu05}. In this way we estimate that effective membrane speeds of $20,000$ m/s are achievable.

A very important conceptual and practical difference between the `membrane-in-the-middle' setup considered here and the original Davis-Fulling-DeWitt moving mirror proposals, as well as Moore's moving cavity end mirror, is that in the latter cases a perfectly reflective mirror moves, whereas in the former case a dielectric membrane of finite reflectivity moves. While a perfect mirror is in any case an idealization, when it moves it leads to pathologies in the theory as recognized by Moore \cite{Moore}.  According to Barton and Eberlein \cite{Barton93} ``In essence, the displacement of a perfectly reflecting surface forces the description of the quantized field out of the original Hilbert space and into another.'' In other words, the creation and annihilation operators for the field for two different mirror positions cannot be defined in the same Hilbert space. Ways around this problem include only working with dielectrics with finite refractive indices, like in references \cite{Barton93} and \cite{Salamone94}, or to use Law's \cite{Law1,Law2,Law3} effective Hamiltonian approach which does not even attempt to describe the true interactions between the field and the charges and currents in the mirror but rather imposes the zero boundary condition at the mirror by hand and then works with the photon operators associated with the continuously evolving `instantaneous' modes. In this paper we adopt a hybrid approach in which the only moving element is a dielectric with a finite refractive index plus we impose the zero boundary condition at the stationary end mirrors.

The plan for this paper is as follows.  In Section \ref{sec:waveequation} we derive the wave equation obeyed by the electric field in the presence of a moving dielectric in the non-relativistic regime. In Section \ref{sec:setup} we solve it for its normal modes in the case of a static membrane.  Section~\ref{sec:adiabaticbasisEOM} considers the case where the membrane is moving and derives the general equations of motion for the optical field by expanding it over the instantaneous normal modes, i.e.\ an adiabatic basis that continuously evolves. An important special case that plays a central role in this paper is that of two modes interacting at an avoided crossing: in Section \ref{sec:CavityEnergy} we give expressions for the energy of such a dichromatic field. In Section~\ref{sec:adiabatictheorem} we give the results of numerically solving the general equations of motion for a membrane moving at a constant velocity which gives a Landau-Zener type sweep of the field through an avoided crossing. In Section~\ref{sec:LocalModeDynamics} we solve the same problem but in the local (diabatic) basis where approximations can be more readily made (and which we verify numerically). These approximations include neglecting the time-dependence of the mode functions and also reducing the second-order-in-time wave equation to one which is first order. We obtain, in Section~\ref{sec:ApproximationCondition}, an analytic criterion for when this first order reduction is valid in terms of the basic parameters of membrane reflectivity and speed.  Section~\ref{sec:RadiationPressure} gives a physical explanation for the change in energy in the cavity as the membrane moves in terms of the work done by the radiation pressure on the membrane. Throughout this paper we try to compare and contrast the classical description of an optical field with the quantum case. This approach culminates in Section \ref{sec:DCE} where we give a detailed quantum description and connect it to the classical field dynamics. We give our conclusions in Section \ref{sec:conclusion}. We have also included four appendices which contain details excluded from the main text; Appendix \ref{app:NonrelativisticApproximation} discusses relativistic corrections to the wave equation in the presence of a moving dielectric membrane, Appendix \ref{app:initialcondition} derives the rather subtle initial conditions for the field that we use for numerically integrating the equations of motion, Appendix \ref{app:quantumEOM} gives the derivation of the quantum equations of motion and Appendix \ref{app:analyticexpressions} sketches the calculation of the various coefficients that appear in the quantum Hamiltonian, as well as giving estimates of their magnitudes in contemporary experiments.

\section{Wave equation in the presence of a moving dielectric}
\label{sec:waveequation}

In their most general form the four Maxwell equations read
\begin{eqnarray}
\nabla \cdot \mathbf{D} & = & \rho_{f} \\
\nabla \cdot \mathbf{B} & = & 0 \\
\nabla \times \mathbf{E} & = & - \frac{\partial \mathbf{B}}{\partial t} \label{eq:faraday} \\
\nabla \times \mathbf{H} & = & \mathbf{J}_{f} + \frac{\partial \mathbf{D}}{\partial t} \label{eq:ampere}
\end{eqnarray}
where $\mathbf{D}$ is the displacement field, $\mathbf{E}$ is the electric field, $\mathbf{H}$ is the magnetizing field and $\mathbf{B}$ is the magnetic field. $\rho_{f}$ and $\mathbf{J}_{f}$ are the free charge and free current, respectively, which exist in the end mirrors but not in the dielectric which is assumed to only contain bound charge and polarization current.  In this paper we follow Law's \cite{Law1,Law2,Law3} approach, where the electromagnetic field is set to zero at the surfaces of the end mirrors by hand. This \emph{effective} theory avoids us having to deal explicitly with the complicated interaction between the fields and  $\mathbf{J}_{f}$ and $\rho_{f}$ in the end mirrors, and we can therefore set these source terms to zero everywhere. We also assume that the electromagnetic properties of the dielectric are linear and isotropic so that they obey the constitutive relations $\mathbf{D}=\epsilon \mathbf{E}$ and $\mathbf{B}=\mu \mathbf{H}$. In fact, we will only consider the case of a non-magnetic dielectric and hence $\mu(x,t) \rightarrow \mu_{0}$, where $\mu_{0}$ is the permeability of free space. Substituting $\mathbf{B}=\mu_{0} \mathbf{H}$ into Faraday's law Eq.\ (\ref{eq:faraday}), taking the curl of both sides, then taking the time derivative of Amp\`{e}re's law Eq.\ (\ref{eq:ampere}), and combining  the two equations gives $-\nabla \times \left( \nabla \times \mathbf{E} \right) = \mu_{0} \partial^2  \mathbf{D} / \partial t^2$. We can use the standard vector identity $\nabla \times \nabla \times \mathbf{E} = \nabla (\nabla \cdot \mathbf{E})-\nabla^2 \mathbf{E}$ to re-write the left hand side, but 
unlike the vacuum case, we do not have $\nabla \cdot  \mathbf{E} = 0$ everywhere. Rather, because $\rho_{f}=0$, we have $\nabla \cdot (\epsilon \mathbf{E})=0$ and hence  $\epsilon (\nabla \cdot \mathbf{E})+ \mathbf{E} \cdot \nabla \epsilon=0$. Thus, the electric field satisfies  
\begin{eqnarray}
\nabla^2 \mathbf{E} + \nabla \left( \mathbf{E} \cdot \nabla \log \epsilon \right)  = \mu_{0} \frac{\partial^2 \epsilon \mathbf{E}}{\partial t^2} .
\label{eq:firstwaveequation}
\end{eqnarray} 
In fact, the second term on the left hand side vanishes identically in the situations we shall consider in this paper where the dielectric function only varies along the cavity axis, whereas the electric field is polarized transversally to this. Because we shall only consider a single polarization we are in essence using the scalar field model \cite{Barton93,Salamone94,Calucci92} in one dimension.

In order to analyse the right hand side of Eq.\ (\ref{eq:firstwaveequation}) we need a model for the dielectric function of a moving membrane. If we assume a gaussian profile of the form 
\begin{equation}
\epsilon_{\mathrm{membrane}}=\alpha \epsilon_{0} \frac{\exp[-(x-vt)^2/w^2]}{\sqrt{\pi} w}
\end{equation}
where $\epsilon_{0}$ is the permittivity of free space and $w$, $v$ and $\alpha$ characterise the membrane's thickness, velocity and dielectric strength, respectively,  we find that the rate of change of the dielectric properties obey
\begin{equation}
\frac{\partial \epsilon_{\mathrm{membrane}}}{\partial t} \le \alpha \epsilon_{0} \sqrt{\frac{2}{\pi}} \frac{v}{w^2} \exp[-1/2]
\label{eq:depsilondt}
\end{equation}  
where  the right hand side has been evaluated at the point $x-vt=w/\sqrt{2}$ where the gaussian changes most rapidly. For the `velocity' we put  $v=5000$ m/s, which is a typical value used in this paper (although velocities up to $v=20,000$ m/s are considered), and guided by the Yale experiments \cite{HarrisJayich,HarrisZwickl} we set $w=50$ nm. In order to estimate the membrane's reflectivity $R$, and hence the value of $\alpha$, we note that when the membrane is much thinner than the wavelength of light, as is the case when $w=50$ nm, we can let $w \rightarrow 0$. In this limit the gaussian reduces to a $\delta$-function and
\begin{equation}
R=\frac{k^2 \alpha^2}{4+k^2 \alpha^2}  .
\label{eq:reflectivity}
\end{equation}
For example, taking $\lambda=2 \pi /k = 785 $ nm and $\alpha= 1.7 \times 10^{-6}$ m gives a membrane reflectivity of 98\%. In fact the $\delta$-function approximation can also be used for thicker membranes provided resonances are avoided where a significant amount of electromagnetic energy is concentrated inside the dielectric \cite{NickPaper}. Inserting the above numbers into Eq.\ (\ref{eq:depsilondt}) gives the estimate $\partial \epsilon_{\mathrm{membrane}}/ \partial t \lesssim 10^{12} \epsilon_{0}$. The factor $10^{12}$ s$^{-1}$ should be compared with the optical frequency  $\omega_{\mathrm{optical}}= \mathcal{O}[10^{15}]$ s$^{-1}$ which characterizes the time-dependence of the electric field. This means that one can reasonably ignore the time derivatives of $\epsilon$ on the right hand side of Eq.\ (\ref{eq:firstwaveequation}) and adopt the standard wave equation but with a space- and time-dependent dielectric function: 
\begin{eqnarray}
\nabla^2 \mathbf{E}   -  \mu_{0} \epsilon(x,t) \frac{\partial^2  \mathbf{E}}{\partial t^2} =0 . 
\label{eq:timedependentMaxwellwave}
\end{eqnarray} 
Inside a dielectric light does not travel at the same speed as in vacuum and this means that the above equation is not relativistically invariant and hence is subject to relativistic corrections when the membrane moves \cite{PiwnickiLeonhardt}.  However,  as shown in Appendix~\ref{app:NonrelativisticApproximation}, these  corrections turn out to be small for the speeds  we consider here and will be neglected. In fact, because we model the dielectric by a $\delta$-function, strictly speaking there is no light inside the medium and the membrane only acts as a boundary condition, somewhat like that due to the end mirrors. It can then be argued that any relativistic corrections due to the medium vanish identically.

\section{Static membrane}
\label{sec:setup}

Our treatment of the dynamics of light in a double cavity is based upon finding the normal modes of the field, and these depend on the position of the membrane. While normal modes only exist for a stationary membrane, which is the focus of the present section, when interpreted as the instantaneous modes at each position of the membrane they can be used as a complete and orthogonal basis for the moving membrane case to be discussed in subsequent sections.

 As above, the membrane is taken to be a thin piece of dielectric material whose spatial profile is modelled by a $\delta$-function. It can transmit light, in contrast to the two end mirrors, which are assumed to be perfectly reflective. Once an initial optical field is established in the double cavity the external pump fields are presumed to be turned off and losses are neglected.   The dynamics of light in the stationary version of this model were studied by Lang \emph{et al} \cite{Lang73} in 1973 in the context of modelling lasers as open systems. One of the subcavities represented the laser cavity and the other, which was much longer, represented the outside world. More recently, the dynamic version of the model has been used by Linington and Garraway \cite{Linington08,liningtonthesis} to study dissipation control in cavities with moving end mirrors, and Casta\~{n}os and Weder \cite{WederCastanos} have used it to find the classical dynamics of a thin end mirror.  

When choosing a coordinate system it is convenient to pretend that the membrane is always located at the origin and the end mirrors are at $x=-L_1$ and $L_2$. The total length of the cavity is $L=L_1+L_2$ and the distance of the membrane from the center is $\Delta L/2$, where $\Delta L  = L_1 - L_2$ is the difference in length between the two subcavities so that $L_{1/2} = (L \pm \Delta L)/2$.  Thus, we write the dielectric function of the double cavity as
\begin{equation}
\epsilon(x,\Delta L)=
\begin{cases}
\epsilon_{0}[1+\alpha\delta(x)], & -L_1<x<L_2 \\
\infty, & x>L_2,\ x<-L_1 .
\end{cases}
\end{equation}
We emphasize that despite this choice of coordinate system, the physical situation we are describing is one in which the membrane is mobile and the end mirrors are fixed.

\begin{figure}[tp]
\includegraphics[width=1\columnwidth]{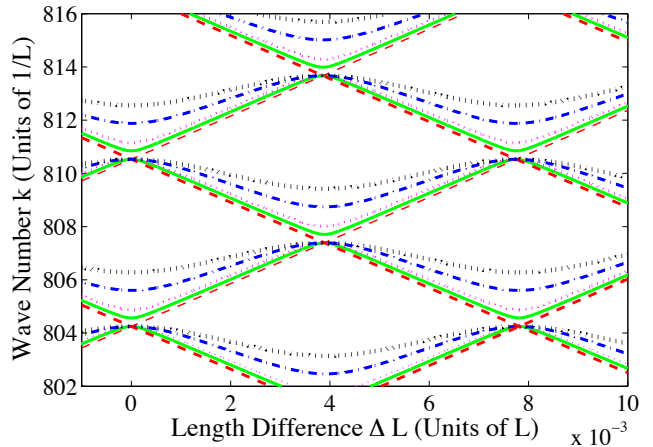}
\caption{The wavenumbers of the normal modes inside a double cavity form a network of avoided crossings when plotted as a function of the difference in length between the two subcavities. The total length of the double cavity is $L=100 \mu$m. The red dashed lines correspond to a perfectly reflective central membrane ($\alpha \rightarrow \infty$). The green solid lines correspond to a membrane of reflectivity $98 \%$ (i.e.\ $\alpha= 1.7 \times10^{-6}$ m), the magenta, small dotted lines correspond to a membrane reflectivity of $91 \%$ (i.e.\ $\alpha=8.0 \times 10^{-7}$m), the blue dashed dotted lines correspond to $61 \%$ (i.e.\ $\alpha=3.1 \times 10^{-7}$m), and the larger, black dotted lines correspond to a membrane reflectivity of $28 \%$ (i.e.\ $\alpha=1.6 \times 10^{-7}$m). All curves except the red curve have avoided crossings. The gap at the avoided crossing ($2\Delta$) goes down as the reflectivity is increased.}
\label{fig:AvoidedCrossing}
\end{figure}

We take the mirrors to lie in the y-z plane and to be translatable along the x-axis,  and consider the case where the electric and magnetic fields are polarized along the $z$ and $y$ axes, respectively. In terms of the vector potential $\mathbf{A}=A(x,t)\hat{\mathbf{z}}$, we have $\mathbf{E}(x,t)=E(x,t) \hat{\mathbf{z}}=-(\di_t A)\hat{\mathbf{z}}$ and $\mathbf{B}(x,t)=B(x,t)\hat{\mathbf{y}}=-(\di_x A)\hat{\mathbf{y}}$. The Maxwell wave equation then takes the form
\begin{equation}
  \frac{\di^{2}E(x,t)}{\di x^{2}}-\mu_{0}\epsilon(x,\Delta L)\frac{\di^{2}E(x,t)}{\di t^{2}}=0.
  \label{eq:StationaryMaxwell}
\end{equation}
The method for solving this equation in terms of normal modes is well known. However, we shall go through it carefully here as a reference for the moving membrane case we tackle in the rest of this paper. To this end we perform a separation of variables, by putting $E(x,t)=C(t)U(x)$, which gives the two equations
\begin{eqnarray}
 && \frac{d^{2}U}{d x^{2}}+k^2 \frac{\epsilon(x, \Delta L)}{\epsilon_{0}} U=0
  \label{eq:Utimeindependent} \\
  && \frac{d^{2}C}{d t^{2}}+\omega^2 C=0 \label{eq:classicalSHO}
\end{eqnarray}
where $\omega^2= c^2 k^2$ is the separation constant and $c=1/\sqrt{\epsilon_{0} \mu_{0}}$ is the speed of light in vacuum. The solutions to Eq.\ (\ref{eq:Utimeindependent}) that obey the boundary conditions $E(x=-L_{1},t)=E(x=L_{2},t)=0$ due to the end mirrors are the global modes of the entire double cavity 
\begin{align}
& U_m(x,\Delta L) = \label{eq:globalmodes} \\
& \begin{cases}
A_{m}(\Delta L)\sin[k_{m}(\Delta L)(x+L_{1}(\Delta L))], & -L_{1}\le x\le0 \\
B_{m}(\Delta L)\sin[k_{m}(\Delta L)(x-L_{2}(\Delta L))], & 0\le x\le L_{2} .  \nonumber 
\end{cases}
\end{align}
The allowed wavenumbers $k_{m}$ satisfy \cite{Lang73}
\begin{equation}
\cos(2k_{m}\Delta L)-\cos(k_{m}L)=2\frac{\sin(k_{m}L)}{\alpha k_{m}} \, 
\label{eq:evaleqn}
\end{equation}
where $m$ is an integer that labels them. Both $k_m$ and $U_m$ depend parametrically on $\Delta L$; when Eq.\ (\ref{eq:evaleqn}) is solved as a function of membrane displacement the result is a network of avoided crossings as shown in Fig.~\ref{fig:AvoidedCrossing}. An important property of the mode functions is that they are orthogonal in the Sturm-Liouville sense. If in addition we  impose normalization they obey
\begin{equation}
  \frac{1}{\epsilon_{0}}\int_{-L_{1}}^{L_{2}} \, \epsilon(x,\Delta L) \, U_{l}(x,\Delta L) \, U_{m}(x,\Delta L) \, \mathrm{dx}=\delta_{lm}. 
  \label{eq:orthonormal}
\end{equation}

The time dependence of the field is determined by Eq.\ (\ref{eq:classicalSHO}) which is the equation of motion for a harmonic oscillator. Factorizing it as $(-i \partial_{t}-\omega_{m})(i\partial_t-\omega_{m})C_{m}=0$ we see that there are two solutions of the form $C_{m}^{\pm}(t) = c_{\pm} \exp[\pm i \omega_{m} t]$. The electric field is a linear combination of $C_{m}^{\pm}(t)$ and must be real. We can therefore put
\begin{equation}
E(x,t)= \sum_{m} \left[ C_{m}^{+}(t) + C_{m}^{-}(t) \right] \, U_m(x,\Delta L) \ ,
\label{eq:staticEfield}
\end{equation}
where the constants $c_{\pm}$ are complex conjugates of each other. Thus, $C_{m}^{+}(t)=[C_{m}^{-}(t)]^{\ast}$, and in this sense the harmonic oscillator equation can be replaced by the single first order equation $(i\partial_t-\omega_{m})C_{m}=0$. Although the harmonic oscillator equation is second order, and hence its solution requires two arbitrary constants (an amplitude and a phase), there is no loss of information in going over to a first order equation because $C_{m}$ is now a complex number specified by two real numbers.
 In the quantum theory $C_{m}(t)$ and $C_{m}^{\ast}(t)$ become lowering and raising operators, respectively, that obey the Heisenberg equations of motion:
\begin{eqnarray}
\left(i\frac{\partial }{\partial t}- \omega_{m} \right)  \hat{C}_{m}(t) & = & 0 \\
\left( -i\frac{\partial}{\partial t} - \omega_{m} \right)  \hat{C}_{m}^{\dag}(t) & = & 0 .
\end{eqnarray}
These equations are not independent: there is really a single equation and its hermitian conjugate. The electric field in Eq.\ (\ref{eq:staticEfield}) becomes an operator proportional to $\hat{C}_{m}^{\dag}(t) + \hat{C}_{m}(t)$ which can be recognized as the position operator (up to constant factors). It is of considerable significance that the quantum equations of motion (and also the classical ones in this case) are first order in time as this ensures that the commutator $[\hat{C}_{m}(t), \hat{C}_{n}^{\dag}(t)]=\delta_{mn}$ is preserved under the dynamics.  This quantization procedure becomes problematic in the presence of a moving membrane because then the dielectric function depends on time and prevents a separation of variables i.e.\ there are no normal modes. Quantization in this situation will be discussed in Section \ref{sec:DCE}.

 In this paper we focus on the dynamics near an avoided crossing, and hence parameterize the two relevant eigenfrequencies as 
\begin{equation}
\omega_{2/1}(\Delta L)=\omega_{\mathrm{av}}\pm\sqrt{\Delta^{2}+\Gamma^2  (\Delta L)}
\label{eq:EvenOddfrequencies}
\end{equation}
where $\Delta$ is half the separation between the two frequencies at the avoided crossing, $\omega_\mathrm{av}$ is their average, and $\Gamma \equiv\sqrt{\gamma} \, \Delta L$ varies linearly with the membrane's displacement from the avoided crossing. In \cite{NickPaper} we showed that for the $\delta$-function membrane model [and for optical frequencies where $\omega = \mathcal{O} (10^{15})$ s$^{-1}$] that
\begin{eqnarray}
   \omega_{0} & \equiv & \frac{2cn\pi}{L} = \omega_{\mathrm{av}} - \Delta \approx \omega_{\mathrm{av}} \\
 \Delta & = & \frac{\omega_{0}}{2}\frac{1}{1+\frac{\omega_{0}^{2}L\alpha}{4c^{2}}} \approx \frac{2 c^2}{\omega_{0} L \alpha} \label{eq:deltadef} \\
  \gamma & = & \frac{\alpha \,  \Delta \, \omega_{0}^{3}}{2Lc^{2}} \approx \frac{\omega_{0}^2}{L^2} \label{eq:gammadef}  
\end{eqnarray}
where $n$ denotes the $n^\mathrm{th}$  pair of modes as counted up from the fundamental mode in a cavity with a perfectly centered and perfectly reflective membrane. For a chosen avoided crossing, the mode corresponding to the lower branch is labelled by the subscript $1$, while that forming the upper branch is labelled by the subscript $2$. When the mirror is perfectly centered the electric field mode functions are either symmetric or antisymmetric. The antisymmetric modes correspond to the lower eigenfrequency ($\omega_1$) of the avoided crossing, while the symmetric state corresponds to the higher eigenfrequency ($\omega_2$). This is in contrast \cite{NickPaper} to the case of material particles governed by the Schr\"{o}dinger equation where the scenario is reversed, i.e.\ the state with the lower eigenvalue is symmetric. 

An alternative basis to the global modes is provided by the local modes 
\begin{equation}
\begin{array}{c}
\phi_{L}(x,\Delta L)=-\sin \theta \ U_{2}(x,\Delta L)+\cos \theta \ U_{1}(x,\Delta L)\\
\phi_{R}(x,\Delta L)=\cos \theta \ U_{2}(x,\Delta L)+\sin \theta \ U_{1}(x,\Delta L)
\end{array}
\label{eq:localmodes}
\end{equation}
where
\begin{equation}
  \sin\theta=-\sqrt{\frac{1}{2}-\frac{\Gamma(\Delta L)}{2\sqrt{\Delta^{2}+\Gamma(\Delta L)^{2}}}}
  \label{eq:sindefn}
\end{equation}
and 
\begin{equation}
  \cos\theta=\sqrt{\frac{1}{2}+\frac{\Gamma(\Delta L)}{2\sqrt{\Delta^{2}+\Gamma(\Delta L)^{2}}}} \ , 
   \label{eq:cosdefn}
\end{equation}
see Appendix D of \cite{NickPaper} for a derivation.  The local modes are localized in the left ($\phi_{L}$) and right ($\phi_{R}$) subcavities.  Although this localization is not perfect, it becomes strong even for moderate membrane reflectivities. The orthonormality of the global modes is inherited by the local modes so that
\begin{equation}
\frac{1}{\epsilon_{0}}\int_{-L_{1}}^{L_{2}}\epsilon(x,\Delta L)\phi_{i}(x,\Delta L)\phi_{j}(x,\Delta L)\mathrm{dx}=\delta_{ij}.
\end{equation}
where $\{i,j\}=\{L,R\}$.
The usefulness of the local basis, when used for dynamics near an avoided crossing, will become apparent in Section~\ref{sec:LocalModeDynamics}. From hence forth the global basis will be referred to as the \emph{adiabatic} basis and the local basis as the \emph{diabatic} basis. This terminology is borrowed from the Landau-Zener problem where the energies of the diabatic states cross linearly as a function time whereas the adiabatic states have an avoided crossing with a minimum gap of $2 \Delta$.  The differences between the diabatic and the adiabatic modes are most stark at the (avoided) crossing; far from the (avoided) crossing they become equal to each other. One note of caution: as explained in Appendix D in reference \cite{NickPaper} the diabatic modes are \emph{not} the same as the perfectly uncoupled modes when the two sides of the cavity are independent except in the limit $\alpha \rightarrow \infty$.

\section{Moving membrane}
\label{sec:adiabaticbasisEOM}

In this section we derive the equations of motion describing the time evolution of light in a double cavity with a moving membrane. Following Linington \cite{liningtonthesis}, we write the evolving electric field in the instantaneous eigenbasis (adiabatic basis) and find differential equations that are second order in time for the corresponding amplitudes. These equations of motion, given in Eq.\ (\ref{eq:LiningtonEquation}) below, will be referred to as the \emph{adiabatic second order equations} (ASOE) and provide us with the most accurate description of the dynamics (they do not assume any adiabatic approximation). The results predicted by the ASOE are the benchmark against which we compare the validity of the approximate dynamics given by the \emph{diabatic second order equations} (DSOE) and the \emph{diabatic first order equations} (DFOE) which will be introduced later.

The adiabatic modes for any instantaneous position of the membrane form a complete basis and we can expand the electric field in terms of them
\begin{equation}
E(x,t)= \sum_{n} c_{n}(t)\exp \left\{ -i\int^t_{t_0}\omega_n(t^{\prime})\mathrm{dt^{\prime}} \right\} U_n(x,t)
\label{eq:globalmodeansatz}
\end{equation}
where the instantaneous mode functions $U_n(x,t)$ at time $t$ are specified in Eq.\ (\ref{eq:globalmodes})  and the time-dependent coefficients $c_{n}(t)$ are in general complex numbers. Although we have not made it explicit, it is understood that the physical electric field is given by the real part of Eq.\ (\ref{eq:globalmodeansatz}). Substituting equation (\ref{eq:globalmodeansatz}) into (\ref{eq:timedependentMaxwellwave}), one finds \cite{liningtonthesis}
\begin{multline}
\underset{n}{\sum} \left[\underbrace{-2i\omega_n \frac{\di}{\di t} (c_n(t)U_n(x,t))}_\mathrm{1} + \underbrace{\frac{\di^2}{\di t^2} (c_n(t)U_n(x,t))}_\textrm{2} \right. \\ \left. \underbrace{-i\frac{\di \omega_n(t)}{\di t}c_n(t)U_n(x,t)}_\textrm{3}\right]\exp\left[-i\int_{t_0}^t\omega_n(t^\prime)\mathrm{dt}^\prime\right] = 0.
\label{eq:LiningtonEquation0}
\end{multline}
Term 1 is by far the dominant one due to the very large optical frequency prefactor. In the slow membrane regime term 2 is small while term 3 is much smaller still because the adiabatic mode can change more significantly in comparison to the rate of change of the optical frequency near an avoided crossing. Right at the avoided crossing, the frequencies are at a maximum or a minimum and hence their rate of change is zero. The relative magnitude of all these terms is analyzed in greater detail in reference \cite{liningtonthesis}. In particular, for faster membrane speeds terms 2 and 3 can become of similar magnitude. 

By projecting out the $m^{\mathrm{th}}$ amplitude using the orthnormality of adiabatic modes, we find from Eq.\ (\ref{eq:LiningtonEquation0}) that the amplitudes corresponding to the adiabatic basis satisfy the \emph{ASOE} \cite{liningtonthesis}
\begin{align}
\ddot{c}_{m}(t)- & i\dot\omega_{m}(t)c_{m}(t)-2i\omega_{m}(t)\dot{c}_{m}(t)
+\underset{n}{\sum}\left\{ [2\dot{c}_{n}(t)- \right. \nonumber \\
& \left. 2i\omega_{n}(t)c_{n}(t)]P_{mn}(t)+c_{n}(t)Q_{mn}(t)\right\}=0 .
\label{eq:LiningtonEquation}
\end{align}
In these equations 
\begin{eqnarray}
\theta_{mn}(t) & \equiv & \int_{t_{0}}^{t}[\omega_{m}(t^{\prime})-\omega_{n}(t^{\prime})]\,\mathrm{dt}^{\prime} \nonumber \\
P_{mn}(t) & \equiv & e^{i\theta_{mn}(t)} \int_{-L_{1}}^{L_{2}}\frac{\epsilon(x,t)}{\epsilon_{0}}U_{m}(x,t)\frac{\partial U_{n}(x,t)}{\partial t} \,\mathrm{dx}
\label{eq:Pdefinition} \nonumber \\
Q_{mn}(t) &  \equiv & e^{i\theta_{mn}(t)}\int_{-L_{1}}^{L_{2}}\frac{\epsilon(x,t)}{ \epsilon_{0}}U_{m}(x,t)\frac{\partial^{2}U_{n}(x,t)}{\partial t^{2}} \,\mathrm{dx}. \nonumber
\end{eqnarray}
The integrals $P_{mn}(t)$ and $Q_{mn}(t)$ depend on the motion of the membrane through the time-dependence of the adiabatic mode functions $U_n(x,t)$. If the membrane is stationary $P_{mn}$ and $Q_{mn}$ vanish and there is no coupling between the different adiabatic modes. 

The coupled differential equations Eq.\  (\ref{eq:LiningtonEquation}) are  second order in time and we therefore need to specify two conditions at the initial time $t_{0}$ in order to solve them.  We choose $c_m(t_0)$ and $\dot{c}_m(t_0)$. However, while $c_m(t_0)$ can be found for any choice of the initial field configuration by projecting it over the expansion given in Eq.\ (\ref{eq:globalmodeansatz}), it is not so obvious what to choose for $\dot{c}_m(t_0)$.   In particular, if we assume that for times $t<t_{0}$ the membrane is stationary then we show in Appendix \ref{app:initialcondition} that the correct initial condition for the time derivatives of the coefficients is $\dot{c}_m(t_0)=-\sum_{n} P_{mn}(t_0)c_n(t_0)$.

\section{Energy of a dichromatic field}
\label{sec:CavityEnergy}

A key quantity in our analysis of the dynamics is the instantaneous energy of the electromagnetic field 
\begin{align}
\mathcal{E} & =\frac{1}{2}\int_{\mathcal{V}}\left[\epsilon(x,t)|E(x,t)|^{2}+\mu_{0}|H(x,t)|^{2}\right]\mathrm{dV} \nonumber \\
& =\frac{\mathcal{A}}{2}\int\left[\epsilon(x,t)|E(x,t)|^{2}+\mu_{0}|H(x,t)|^{2}\right]\mathrm{dx} \label{eq:energygeneral}
\end{align}
where $H(x,t)=B(x,t)/\mu_0$, $\mathcal{A}$ is the area of the mode functions, and $\mathcal{V}$ is the volume of the cavity. Note that the vanishing volume of the $\delta$-function membrane means that there is no contribution from it.  In this paper we are interested in the field dynamics when passing through an avoided crossing where attention can be restricted to just two modes. We therefore consider a dichromatic field in the adiabatic basis with frequencies $\omega_{1}$ and $\omega_{2}$. The total electric field can then be written as 
\begin{multline}
E(x,t)=c_{1}(t)\exp[-i\theta_{1}(t)]U_{1}(x,t) + \\ c_{2}(t)\exp[-i\theta_{2}(t)]U_{2}(x,t)
\end{multline}
where $U_m(x,t)$ is defined in Eq.\ (\ref{eq:globalmodes}) and
\[
\theta_{m}(t)=\int_{t_{0}}^{t}\omega_{m}(t^{\prime})\mathrm{dt}^{\prime}.
\]
Hence, the energy per unit area becomes
\[
\frac{\mathcal{E}}{\mathcal{A}}=\frac{\epsilon_{0}}{2}\left\{ |c_{1}(t)|^{2}+|c_{2}(t)|^{2}\right\} +\frac{\mu_{0}}{2}\int_{-L_{1}}^{L_{2}}|H(x,t)|^{2}\mathrm{dx}.
\]
Assuming that, as usual, the magnetic field makes a contribution to the energy equal to that of the electric field, we  arrive at the following expression for the total energy per unit area:
\begin{equation}
\frac{\mathcal{E}}{\mathcal{A}}=\epsilon_{0}\left\{ |c_{1}(t)|^{2}+|c_{2}(t)|^{2}\right\} .
\label{eq:lightenergy}
\end{equation}

In time-independent situations the Hamiltonian gives the energy of a system. However, this is not necessarily true in time-dependent systems where the Hamiltonian still plays the role of the generator of dynamics but need not coincide with the energy.  Reference \cite{schutzhold98} proves that  Eq.\ (\ref{eq:energygeneral}) is the correct expression for the instantaneous energy even in time-dependent situations. Although our approach to finding the dynamics in this paper is based upon the wave equation rather than the Hamiltonian,  we shall have occasion to derive the Hamiltonian in Sec.\ \ref{sec:DCE} and will find that it contains extra velocity-dependent terms not present in Eq.\ (\ref{eq:energygeneral}).

\section{Field dynamics while traversing an avoided crossing}
\label{sec:adiabatictheorem}

In this section we apply the full ASOE derived in Section \ref{sec:adiabaticbasisEOM} to the case of a Landau-Zener style sweep of the membrane through an avoided crossing.   The Landau-Zener problem is a rare example of where the time-dependent Schr\"{o}dinger equation can be solved exactly and the adiabaticity of the motion evaluated analytically.   The Schr\"{o}dinger equation for the Landau-Zener problem is
\begin{align}
i \frac{d}{dt} & \left(
\begin{array}{c}
a_{L}\\
a_{R}
\end{array}
\right)  =  H_{\mathrm{LZ}}(t)
 \left(
\begin{array}{c}
a_{L}\\
a_{R}
\end{array}
\right) \label{eq:SchrodLZ}
\end{align}
where, in the notation used in this paper, the Landau-Zener Hamiltonian takes the form
\begin{equation}
H_{\mathrm{LZ}}(t) \equiv  \left( 
\begin{matrix}
  \omega_{\mathrm{av}}+ \Gamma(t) & \Delta \\
 \Delta &  \omega_{\mathrm{av}}- \Gamma(t)
\end{matrix} 
\right) . \label{eq:HLZ}
\end{equation} 
It describes the case where two \emph{diabatic} levels cross linearly in time and in the double cavity system this corresponds to a membrane moving at constant velocity $v$. Given that the membrane displacement is $\Delta L/2$, and that $\Gamma(t) \equiv \sqrt{\gamma} \, \Delta L(t)$ [see  Eq.\ (\ref{eq:EvenOddfrequencies}) and the definitions given below it],  for a Landau-Zener sweep we must put 
\begin{equation}
\Gamma(t)=\sqrt{\gamma} \, \Delta L(t)= \sqrt{\gamma} 2 v t.
\end{equation}
 The diabatic levels cross at $t=0$ and have a constant coupling given by $\Delta$. In the \emph{adiabatic} basis the same Schr\"{o}dinger equation becomes
\begin{align}
i \frac{d}{dt} & \left(
\begin{array}{c}
c_{2}\\
c_{1}
\end{array}
\right)  =   \left( 
\begin{matrix}
  \omega_{2}(t) & 0 \\
 0 &  \omega_{1}(t)
\end{matrix} 
\right)
 \left(
\begin{array}{c}
c_{2}\\
c_{1}
\end{array}
\right)
\end{align}
where $\omega_{2/1}(t)=\omega_{\mathrm{av}} \pm \sqrt{\Delta^2+\Gamma^{2}(t)}$. There is an avoided crossing between the two adiabatic states with a gap of $2 \Delta$  at $t=0$.  If the system starts in one of the adiabatic states at $t=-\infty$ the probability that it has made a transition to the other adiabatic state by $t=+\infty$ is given by \cite{Landau,Zener,Stenholm}
\begin{equation}
P_{\mathrm{LZ}}=\exp \left[-  \pi \Delta^2/ (2 v \sqrt{\gamma}) \right].
\label{eq:P_LZ}
\end{equation}
The process becomes more adiabatic as the velocity $v$ is reduced; the population transfer  approaches zero exponentially fast in $1/v$.

 \begin{figure}[tp]
 \includegraphics[width=\linewidth]{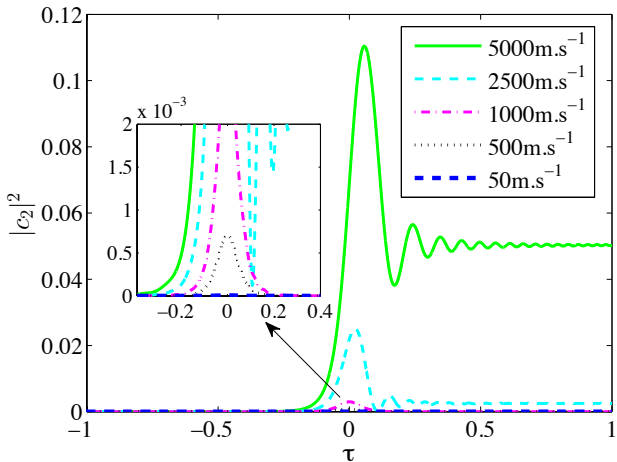}
 \caption{Dynamics of an initially empty mode when traversing an avoided crossing at five different speeds.  We simulated the field dynamics using the ASOEs given in Eq.\ (\ref{eq:adiabaticLinington}) in the two-level approximation near an avoided crossing. $c_{2}$ is the amplitude associated with the upper adiabatic mode, where the initial condition is $c_1=1$ and $c_2=0$. According to Eq.\ (\ref{eq:lightenergy}), $\vert c_{n} \vert^2$ is proportional to the electromagnetic energy of the $n^{\mathrm{th}}$ mode.  We see that as the membrane speed goes down, the energy pumped into  the initially unpopulated mode tends to zero. Parameters: membrane reflectivity $98\%$ (i.e.\ $\alpha=1.5 \times 10^{-6}$m); length of double cavity $100 \mu$m; maximum membrane displacement $\Delta L/2 = \pm 1 \times 10^{-7}$m.  The adiabatic modes shown are those with $n=128$, where we label the modes in terms of the wavenumbers for a perfectly reflecting membrane for which $k_{n}=2 \pi n /(L\pm\Delta L)$. These perfectly localized modes come in pairs that are degenerate at $\Delta L=0$.  }
 \label{fig:AdiabaticCriteriaSingleMode2}
 \end{figure}
 \begin{figure}[tp]
 \includegraphics[width=\linewidth]{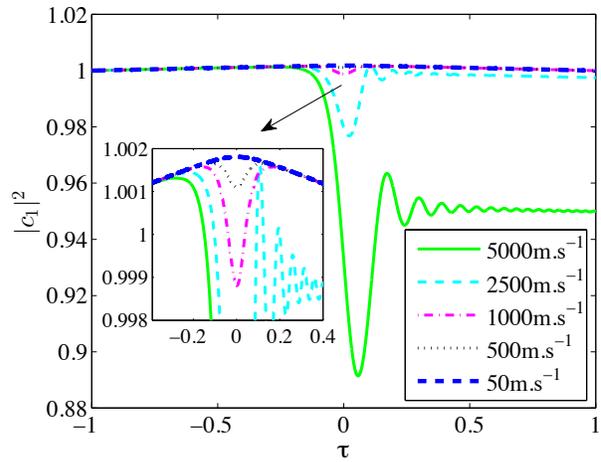}
 \caption{Dynamics of an initially excited mode when traversing an avoided crossing at different speeds as calculated using the ASOEs. This figure is for exactly the same setup as Fig.\ \ref{fig:AdiabaticCriteriaSingleMode2} except here we plot the results for the lower mode. We see that as the membrane speed is decreased the energy of this mode is not conserved but has a slight upward curve. Combined with Fig.\ \ref{fig:AdiabaticCriteriaSingleMode2}, this tells us that whilst the slowly moving membrane limit is sufficient to avoid nonadiabatic transitions, energy is not conserved.}
 \label{fig:AdiabaticCriteriaSingleMode}
 \end{figure}

We should not expect the Landau-Zener theory to apply to the classical electromagnetic field because the latter does not obey the Schr\"{o}dinger equation. Nevertheless, as we shall see, there are regimes where we can map the passage of the electromagnetic field through an avoided crossing onto the Landau-Zener problem.   In particular, we find that decreasing the membrane speed is a sufficient criteria for achieving adiabaticity in the Maxwell wave equation in the sense of vanishing transfer between adiabatic modes. However, contrary to the Schr\"{o}dinger case, we find that even at very slow membrane speeds we do not conserve the sum $\vert c_{1}(t) \vert^2+\vert c_{2}(t) \vert^2$. In quantum mechanics  the coefficients $c_{n}(t)$ are probability amplitudes and the sum of their squares represents the total probability which is conserved under the unitary evolution provided by the Schr\"{o}dinger equation. The same is not true in the Maxwell case where, as we saw in Section \ref{sec:CavityEnergy},  the sum of the squares represents the total energy which is in general not conserved when an external parameter is varied. Physically, the electromagnetic field interacts with the membrane via radiation pressure and as a result energy can be transferred back and forth between the field and the external agent moving the membrane. There is always radiation pressure on the membrane (except right at an avoided crossing) and therefore some energy is pumped into/out of the system regardless of how slowly the membrane is being moved. This is a fundamental difference between adiabaticity in the Schr\"{o}dinger and Maxwell wave equations.

We consider the situation where the membrane moves at constant speed $v$ from position $x=-L_0$ to $L_0$ over the time $t=-T_0$ to $T_0$. The displacement of the membrane from the center is given by $\Delta L/2$, 
\begin{equation}
  \frac{\Delta L(t)}{2}= \frac{L_0}{T_0}t
\end{equation}
and $v=L_0/T_0$. We investigate the effects of varying the speed by fixing $L_0$ and changing $T_0$. It is useful to introduce the scaled time variable
\begin{equation}
  \tau=\frac{t}{T_0}=\lambda t,\: -1\le\tau\le1
\end{equation}
i.e. $\lambda = \frac{1}{T_0}$. In terms of these variables the ASOE given in Eq.\ (\ref{eq:LiningtonEquation}) become
\begin{align}
\frac{dc_{m}}{d\tau} = -\frac{d\omega_{m}}{d\tau}\frac{c_m}{2\omega_{m}}- & \frac{i\lambda}{2\omega_{m}}\frac{d^2 c_{m}}{d\tau^2}-\underset{n}{\sum}\left\{ \left[\frac{i\lambda}{\omega_{m}}\frac{d c_{n}}{d\tau}+\right.\right. \nonumber \\
& \left.\left.\frac{\omega_{n}}{\omega_{m}}c_{n}\right]\bar{P}_{mn}+\frac{i\lambda}{2\omega_{m}}c_{n}\bar{Q}_{mn}\right\}
\label{eq:adiabaticLinington}
\end{align}
where
\begin{eqnarray}
  \bar{\theta}_{mn} & \equiv & \frac{1}{\lambda}\int_{-1}^{\tau}\left[\omega_{m}(\tau^{\prime})-\omega_{n}(\tau^{\prime})\right]d\tau^{\prime} \nonumber \\
  \bar{P}_{mn} & \equiv & e^{i\bar{\theta}_{mn}}  \int_{-L_{1}}^{L_{2}}\frac{\epsilon(\tau,x)}{\epsilon_{0}}U_{m}(\tau,x)\di_{\tau}U_{n}(\tau,x)dx\nonumber \\
  \bar{Q}_{mn} & \equiv & e^{i\bar{\theta}_{mn}} \int_{-L_{1}}^{L_{2}}\frac{\epsilon(\tau,x)}{\epsilon_{0}}U_{m}(\tau,x)\di_{\tau}^{2}U_{n}(\tau,x)dx. \nonumber
\end{eqnarray}

Let us assume that a single mode $c_m$ is initially populated and all other modes are empty. When $T_0 \rightarrow \infty$, we have $\lambda \rightarrow 0$ and the factors $\bar{P}_{mn}$ and $\bar{Q}_{mn}$ for $n \neq m$ approach zero due to the presence of the phase term which oscillates infinitely rapidly in that limit. However, the diagonal terms $\bar{P}_{mm}$ and $\bar{Q}_{mm}$ have no such phase term and are generally non-zero. The term $(d\omega_m/d\tau)(c_m/2\omega_m)$ in Eq.\ (\ref{eq:adiabaticLinington}) is related to the slope of the frequency and is non-zero everywhere except exactly at the avoided crossing. Since the first term and the diagonal part of the fourth term of the right hand side of Eq.~(\ref{eq:adiabaticLinington}) do not approach zero in the slow membrane limit, $d c_m/d\tau$ does not approach zero. Meanwhile, the rates of change of all initially unpopulated states do approach zero in the slow membrane limit because the first and fourth terms in Eq.~(\ref{eq:adiabaticLinington}) depend on the mode population. This indicates that all the initially empty modes continue to remain empty in the slow membrane limit despite the fact that the initially occupied mode can in general change its amplitude no matter how slowly the membrane is moved.

 \begin{figure}[tp]
 \includegraphics[width=\linewidth]{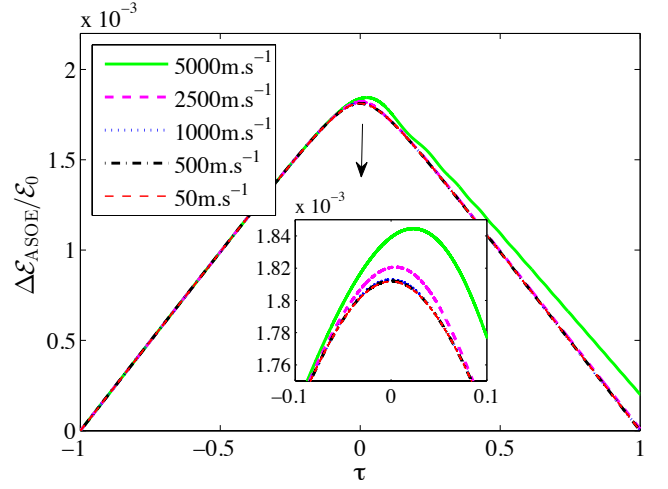}
 \caption{The fractional change in total energy of the system as it traverses an avoided crossing as calculated using the ASOEs. Here $\mathcal{E}_{0}$ is the initial energy and we use the same parameters as in Figs. \ref{fig:AdiabaticCriteriaSingleMode2} and \ref{fig:AdiabaticCriteriaSingleMode}. The plot shows that even at very slow membrane speeds the energy change as a function of time does not vanish but instead tends to a limiting curve. Hence, even though $v \rightarrow 0$, we find $\sum_{n} \vert c_{n}(\tau) \vert^2 \neq \mathrm{constant}$, confirming that adiabaticity does not imply energy conservation.}
 \label{fig:AdiabaticityEnergy}
 \end{figure}

This analysis of the equations of motion is supported by the numerical results shown in Figs.\ \ref{fig:AdiabaticCriteriaSingleMode2},\ref{fig:AdiabaticCriteriaSingleMode}, and \ref{fig:AdiabaticityEnergy} where we plot dynamics for a pair of adiabatic modes as they traverse an avoided crossing at various speeds.  At higher speeds energy is removed from the initially excited mode  and transferred to the initially empty mode as can be seen from the almost perfect mirror symmetry of the $\vert c_{1} \vert^2$ and $\vert c_{1} \vert^2$ curves about the midpoint $\vert c_{1} \vert^2=\vert c_{1} \vert^2=0.5 $. This is the type of behaviour we would expect in the standard Landau-Zener problem with the Schr\"{o}dinger equation. And at very low speeds we see from Fig.\ \ref{fig:AdiabaticCriteriaSingleMode2} that the amplitude of the initially empty mode remains zero indicating adiabatic evolution, as expected. However, in Fig.\ \ref{fig:AdiabaticCriteriaSingleMode} we see that at low speeds the various curves for the initially excited mode converge towards a limiting curve where there is a finite change in energy of the mode.
To make this point clearer we plot the change in total energy of the system in Fig.~\ref{fig:AdiabaticityEnergy}. To be precise, we plot the change in energy divided by the initial energy $|c_1|^2+|c_2|^2-1=\Delta \mathcal{E}_{\mathrm{ASOE}}/\mathcal{E}_0$, and as one can see no matter how slowly the membrane is moved the energy pumped into the system converges to a curve that always lies above the zero axis. We also note that the slow speed limiting curve is symmetric about the avoided crossing at $\tau=0$, indicating that whatever energy is pumped in when approaching the avoided crossing is pumped out as it recedes. However, at higher speeds there is a noticeable net energy gain by the electromagnetic field.

\section{Dynamics in the diabatic basis}
\label{sec:LocalModeDynamics}

\begin{figure}
\includegraphics[width=\linewidth]{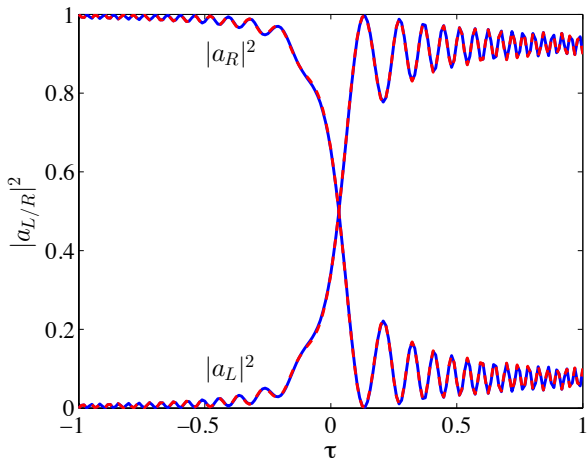}
\caption{Dynamics on traversing an avoided crossing as seen in the diabatic basis, with the field initially localized on the right. Membrane speed: $5000 \ \mathrm{m s}^{-1}$. All other parameters are the same as in Fig.\ \ref{fig:AdiabaticCriteriaSingleMode2}. The results were calculated using both the ASOE and DSOE schemes with the two sets of curves lying right on top of each other.}
\label{fig:FullvsFirstOrder}
\end{figure}

In this Section we first obtain the second-order-in-time equations of motion in the diabatic basis (DSOE) and then approximate them to first-order-in-time equations (DFOE).
So far we have worked in the adiabatic basis which corresponds to the instantaneous normal modes of the double cavity. One feature of this basis is that as an avoided crossing is traversed the two mode functions involved radically change their structure by exchanging the sides upon which they are principally localized, see Fig.\ 4 in \cite{NickPaper}. Conversely, the expansion amplitudes $c_{m}(t)$ in the adiabatic basis experience only exponentially small changes in the slow membrane regime. The opposite is true for the diabatic basis where the mode functions hardly change but there is a large change in the amplitudes. The diabatic basis is advantageous for making analytic calculations because to a good approximation we can ignore the time dependence of the mode functions and focus all our attention on the amplitudes, a fact Zener points out in his original paper \cite{Zener}. We shall confirm this property below. 

 Assuming as before that the membrane motion is restricted to be in the vicinity of an avoided crossing, we employ the two-level approximation and let
\begin{equation}
E(x,t)=a_L(t)\phi_L(x,t)+a_R(t)\phi_R(x,t).
\label{eq:localmodeelectricfield}
\end{equation}
Substitution into the Maxwell wave equation given in Eq.\ (\ref{eq:timedependentMaxwellwave}),  and neglecting the terms $\dot{\phi}_{L/R}$ and $\ddot{\phi}_{L/R}$, yields
\begin{equation}
a_L(t){\phi}^{\prime \prime}_L+a_R(t){\phi}^{\prime \prime}_R=\mu_{0}\epsilon(x,t)[\ddot{a}_L(t)\phi_L+\ddot{a}_R(t)\phi_R]
\label{eq:PartialSecondOrder0}
\end{equation}
where the dots indicate time derivatives and the dashes spatial derivatives.  The diabatic modes are not normal modes of the double cavity and so even for a stationary membrane the light oscillates back and forth between the left and right modes in a fashion analogous to the Rabi oscillations of a two-level atom interacting with a single mode field. The combined effect of this intrinsic oscillation and the moving membrane leads to a much larger rate of change of the diabatic amplitudes compared to the adiabatic amplitudes.

\begin{figure}[t]
\includegraphics[width=\linewidth]{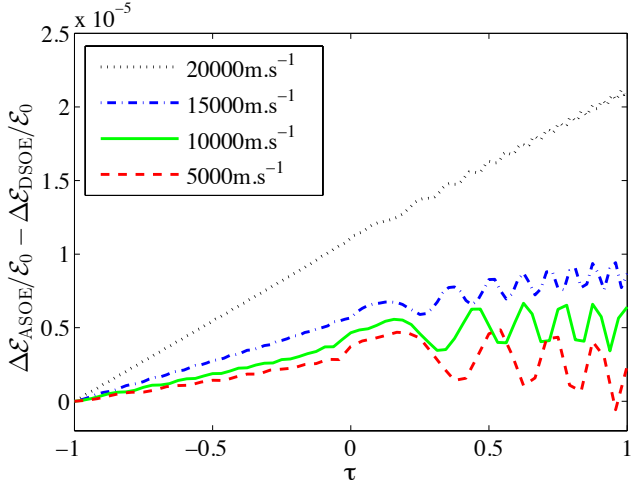}
\caption{A comparison of the fractional change in energy calculated using the ASOE and DSOE for the same avoided crossing dynamics as shown in Fig.\ \ref{fig:FullvsFirstOrder} except that here we also vary the membrane speed. We see that the order of magnitude of difference between ASOE and DSOE is of the order of $1 \times 10^{-5}$. Here, $\Delta \mathcal{E}_{\mathrm{ASOE}}/\mathcal{E}_0$ is generated by Eq.\ (\ref{eq:LiningtonEquation}) and $\Delta \mathcal{E}_{\mathrm{DFOE}}/\mathcal{E}_0$ is generated by Eq.\ (\ref{eq:PartialSecondOrder}). Although a speed of $20,000$ ms$^{-1}$ seems very high, such effective speeds can be achieved by changing the background index of refractions rather than physically moving the mirror.}
\label{fig:FullvsPartialReflectivity98}
\end{figure}
\begin{figure}[t]
\includegraphics[width=\linewidth]{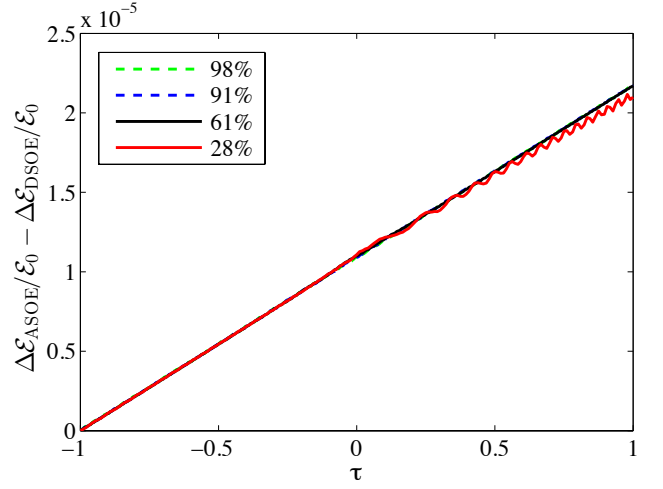}
\caption{A comparison of the fractional change in energy calculated using the ASOE and DSOE for the same avoided crossing dynamics as shown in Fig.\ \ref{fig:FullvsFirstOrder} except that here we also vary the membrane reflectivity. The results lead us to the same conclusion as in Fig.\ \ref{fig:FullvsPartialReflectivity98}. Here, $\Delta \mathcal{E}_{\mathrm{ASOE}}/\mathcal{E}_0$ is generated by equation~(\ref{eq:LiningtonEquation}) and $\Delta \mathcal{E}_{\mathrm{DFOE}}/\mathcal{E}_0$ is generated by equation~(\ref{eq:PartialSecondOrder}).}
\label{fig:FullvsPartialSpeed5000}
\end{figure}

In order to find the spatial derivatives of the diabatic modes in Eq.\ (\ref{eq:PartialSecondOrder0}), we express each diabatic mode in the adiabatic basis whose second derivatives we know in terms of the Sturm-Liouville relationship given in Eq.\ (\ref{eq:Utimeindependent}), $\di_x^2U_m(x,\Delta L)= - (\epsilon(x,\Delta L)  / \epsilon_0) k_m^2 U_m(x,\Delta L)$, and then convert back to the diabatic basis. In matrix form, we find \cite{NickPaper}
\begin{align}
- & \left(
\begin{array}{c}
\ddot{a}_{L}\\
\ddot{a}_{R}
\end{array}
\right)  = \\  & \left( 
\begin{matrix}
  \omega^2_2 \cos^2\theta+\omega_1^2 \sin^2\theta & (\omega_1^2-\omega_2^2) \cos\theta\sin\theta \\
   (\omega_1^2-\omega_2^2) \cos\theta\sin\theta & \omega_1^2 \cos^2\theta+\omega_2^2 \sin^2\theta
\end{matrix} 
\right)  \nonumber 
 \left(
\begin{array}{c}
a_{L}\\
a_{R}
\end{array}
\right)
\end{align}
or
\begin{equation}
\left(
\frac{d^2}{dt^2} +  M_{\mathrm{DSOE}} \right)  
 \left(
\begin{array}{c}a_{L}\\
a_{R}
\end{array}
\right) =0   \label{eq:PartialSecondOrder}
\end{equation}
where
\begin{equation}
M_{\mathrm{DSOE}}= \left(
\begin{matrix}
\left[\omega_{\mathrm{av}}+\Gamma(t)\right]^{2}+\Delta^{2} & 2\Delta \, \omega_{\mathrm{av}} \\
2\Delta \, \omega_{\mathrm{av}} & \left[\omega_{\mathrm{av}}-\Gamma(t)\right]^{2}+\Delta^{2}
\end{matrix}
\right)  \label{eq:DSOEmatrix}
\end{equation}
and we have made use of the identities
\begin{eqnarray}
&& \sin \theta \cos \theta \left( \omega_{1}^{2} - \omega_{2}^2 \right) = 2 \Delta \omega_{\mathrm{av}} \\
&& \omega_{1}^2 \cos^{2} \theta +  \omega_{2}^2 \sin^{2} \theta  = (\omega_{\mathrm{av}}-\Gamma )^2 + \Delta^2 \\
&& \omega_{2}^2 \cos^{2} \theta +  \omega_{1}^2 \sin^{2} \theta  = (\omega_{\mathrm{av}}+\Gamma)^2 + \Delta^2 .
\end{eqnarray}
We refer to Eq.\ (\ref{eq:PartialSecondOrder}) as the \emph{diabatic second order equations} (DSOE).

\begin{figure}[t]
\includegraphics[width=\linewidth]{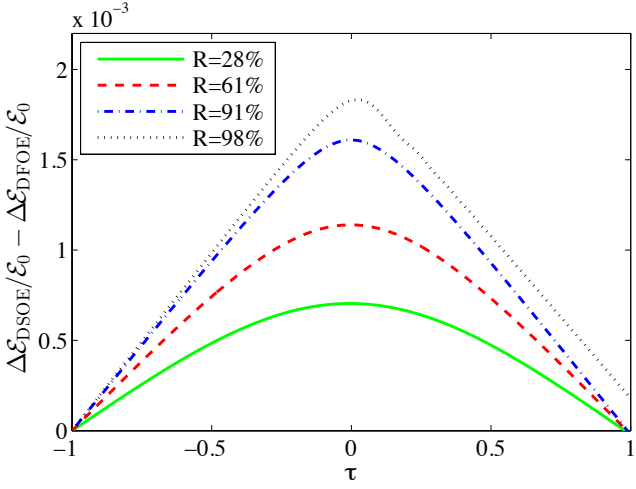}
\caption{This figure shows the trend of agreement between DSOE and DFOE as we vary the membrane reflectivity for the same avoided crossing dynamics as shown in Fig.\ \ref{fig:FullvsFirstOrder}. For first order dynamics, $\Delta \mathcal{E}_{\mathrm{DFOE}}/\mathcal{E}_0$ has to be identically zero, while for second order dynamics it is generally nonzero. Hence the difference of this quantity from zero can be used to quantify the validity of the first order model. As reflectivity goes up, the first order approximation becomes less valid.  }
\label{fig:PartialvsFirstSpeed5000}
\end{figure}

The DSOE are strongly reminiscent of the second order in time harmonic oscillator equation given in Eq.\ (\ref{eq:classicalSHO}) for the static membrane, albeit in the present case there are two modes involved. This begs the question as to whether the DSOE can be factorized and reduced to a first order equation like the harmonic oscillator equation can. To this end we note that $M_{\mathrm{DSOE}}=H_{\mathrm{LZ}}^2$, and thus it is tempting to write Eq.\ (\ref{eq:PartialSecondOrder}) as
\begin{equation}
\left(i\frac{d}{dt} - H_{\mathrm{LZ}}  \right) \left(- i\frac{d}{dt} -H_{\mathrm{LZ}}  \right)  \left( \begin{array}{c}a_{L}\\
a_{R}
\end{array}
\right) =0 \  \mbox{?} \label{eq:DSOEFactorize}
\end{equation}
This factorization is correct in the time-independent case, but due to the time-dependence of $H_{\mathrm{LZ}}$, when the left hand side of Eq.\ (\ref{eq:DSOEFactorize}) is expanded there is an extra term $-i \dot{H}_{\mathrm{LZ}}$ not present in the DSOE given in Eq.\ (\ref{eq:PartialSecondOrder}).

\begin{figure}[t]
\includegraphics[width=\linewidth]{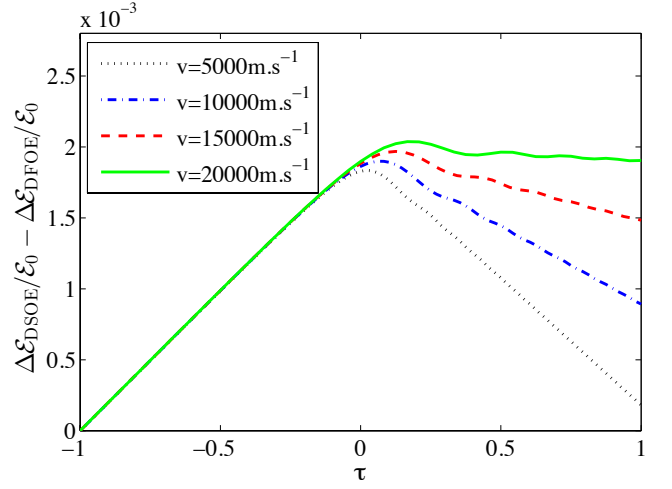}
\caption{This figure shows the trend of agreement between DSOE and DFOE as we vary the membrane speed for the same avoided crossing dynamics as shown in Fig.\ \ref{fig:FullvsFirstOrder}. As our intuition would suggest, the first order approximation becomes less valid for higher speeds. }
\label{fig:PartialvsFirstReflectivity98}
\end{figure}

Although the DSOE cannot be exactly reduced to first-order-in-time equations, a first order approximation can be derived as we now show.  
We start by transforming the left/right mode amplitudes  
\begin{equation}
a_{L/R}=\tilde{a}_{L/R}\exp\left\{ -i\int_{t_{0}}^{t}\beta_{L/R}(t^{\prime})dt^{\prime}\right\}
\label{eq:DFOEansatz}
\end{equation}
where
\begin{equation}
\beta_{L/R}(t) \equiv \sqrt{(\Gamma(t)\pm\omega_{\mathrm{av}})^{2}+\Delta^{2}}.
\label{eq:betadef}
\end{equation}
Substituting for the new variables  removes the fast oscillations 
\begin{align}
\ddot{a}_{L/R}= & \left\{\ddot{\tilde{a}}_{L/R}-2i\beta_{L/R}\dot{\tilde{a}}_{L/R}-i\dot{\beta}_{L/R}\tilde{a}_{L/R}- \right.\nonumber \\ 
& \left. \beta_{L/R}^{2}\tilde{a}_{L/R}\right\} \exp\left\{ -i\int_{t_{0}}^{t}\beta_{L/R}(t^{\prime})dt^{\prime}\right\} .
\label{eq:SecondtoFirstOrder}
\end{align}
Intuitively, we expect that during a slow sweep the first and third terms on the right hand side will be small and hence we shall ignore them. We will check the validity of these assumptions  numerically below (and in Section \ref{sec:ApproximationCondition} we derive an analytic criterion for the validity of the first order approximation). We have that
\begin{equation}
i\dot{\tilde{a}}_{L/R}=\frac{\omega_{\mathrm{av}}\Delta}{\beta_{L/R}}\tilde{a}_{R/L}\exp\left\{ \pm i\int_{t_{0}}^{t}[\beta_{L}-\beta_{R}]dt^{\prime}\right\} .
\end{equation}
Assuming $\omega_{\mathrm{av}}$ is very large we can put
\begin{equation}
\beta_{L/R}(t)\approx\omega_{\mathrm{av}}\left\{ 1\pm\frac{\Gamma(t)}{\omega_{\mathrm{av}}}+\frac{1}{2}\frac{\Delta^{2}}{\omega_{\mathrm{av}}^{2}}\right\}.
\end{equation}
Hence, $\beta_{L}-\beta_{R}\approx2\Gamma$ and $\omega_{\mathrm{av}}/\beta_{L/R}\approx1$, giving
\begin{equation}
i\dot{\tilde{a}}_{L/R}=\Delta \, \tilde{a}_{R/L}\exp\left\{ \pm2i\int_{t_{0}}^{t}\Gamma(t^{\prime})dt^{\prime}\right\}.
\end{equation}
Changing the variables back to $a_{L/R}$, we finally obtain 
\begin{equation}
i\left(\begin{array}{c}
\dot{a}_{L}\\
\dot{a}_{R}
\end{array}\right)=\left(
\begin{matrix}
\omega_{\mathrm{av}}+\Gamma(t) & \Delta \\
\Delta & \omega_{\mathrm{av}}-\Gamma(t) 
\end{matrix}
\right)\left(\begin{array}{c}
a_{L}\\
a_{R}
\end{array}\right) .
\label{eq:PartialFirstOrder}
\end{equation}
In the case that the membrane moves at a constant speed, so that $\Gamma(t)$ is linear in time,  these are exactly the Landau-Zener equations [see Eq. (\ref{eq:SchrodLZ})]. We refer to Eq.\ (\ref{eq:PartialFirstOrder}) as the \emph{diabatic first order equations} (DFOE).

The results of numerically simulating the dynamics using the ASOE and DSOE schemes are compared in Figs.\ \ref{fig:FullvsFirstOrder}, \ref{fig:FullvsPartialReflectivity98}, and \ref{fig:FullvsPartialSpeed5000}. In each case the light is initially located on the right side of the cavity and the membrane is moved from left to right.  When the calculation is done using the ASOE we can still plot the results in the diabatic basis by switching basis using Eq.\ (\ref{eq:localmodes}).  From Fig.~\ref{fig:FullvsFirstOrder} we see that the results using the ASOE and DSOE lie on top of each other so that their differences are small relative to the order of magnitude of the amplitudes themselves. Hence, for these parameters we are safe in ignoring the time-dependence of the diabatic mode functions. To get a closer look at the differences, we compute the change in energy relative to its initial value $\Delta \mathcal{E}/\mathcal{E}_0$ for the two sets of equations of motion.  From Figs.\ \ref{fig:FullvsPartialReflectivity98} and \ref{fig:FullvsPartialSpeed5000}, we see that as long as the membrane motion is close to the avoided crossing, the difference is of the order of $10^{-5}$ even for speeds as high as $20,000$ ms$^{-1}$.

Let us now check the validity of the first-order-in-time approximation embodied in the DFOE approach, i.e.\ how good of an approximation it is to ignore the first and the third terms in Eq.\ (\ref{eq:SecondtoFirstOrder}).   In Figs.~\ref{fig:PartialvsFirstSpeed5000} and \ref{fig:PartialvsFirstReflectivity98} we compare the relative change in energy with time using the DSOE and the DFOE. The difference between the first and second order models is directly related to the energy pumped into and out of the system because the first order dynamics preserves $\sum_{n} \vert c_{n} \vert^2$, meaning that $\Delta \mathcal{E}_{\mathrm{DFOE}}$ is identically zero.  We can see that for increasing reflectivity and speed, the first order approximation becomes less valid. Nevertheless, in the optical frequency regime it is a very good approximation as the discrepancy is only of the order of  $10^{-3}$. This number also shows that ignoring the time dependence of the diabatic mode functions is a much smaller effect than the first order reduction of the DSOE to the DFOE.

The finding that the first-order-in-time approximation becomes less valid as reflectivity is increased appears to be in contradiction to the results in our previous paper \cite{NickPaper}. In particular, Fig.\ 7 of \cite{NickPaper} shows that the approximation becomes better as the coupling $\Delta$ is decreased. This might be interpreted (erroneously) as saying that reflectivity should be increased for a better match. That paper showed that the first order equations of motion depend on a single dimensionless parameter $v \sqrt{\gamma} / \Delta^2$, a result which is consistent with the Landau-Zener transition probability given in Eq.\ (\ref{eq:P_LZ}), whereas the second order equations of motion depend additionally upon the dimensionless quantity $\Delta/\omega_{\mathrm{av}}$. Therefore, a comparison of the two dynamics where $v \sqrt{\gamma} / \Delta^2$ is held constant but $\Delta/ \omega_{\mathrm{av}}$ is varied should agree in the limit $\Delta/ \omega_{\mathrm{av}} \rightarrow 0$. This is correct. However, holding $v \sqrt{\gamma} / \Delta^2$ constant and reducing $\Delta$ implies that the speed $v$ must also be decreasing, ensuring that the first order dynamics becomes a better approximation as higher order time derivatives present in the corrections become smaller. Such a comparison of the dynamics is not a good test of the role  of reflectivity because it is not just the reflectivity that is varied. In this paper, and in particular in Fig.\ \ref{fig:PartialvsFirstSpeed5000}, we study a different situation: we fix the initial and final mirror positions and then sweep through the avoided crossing at a fixed speed while varying the  reflectivity.

\section{Analytic criterion for validity of first order dynamics}
\label{sec:ApproximationCondition}

In the previous section we presented numerical evidence showing that the second order Maxwell wave equation can, in certain regimes, be approximated by a first-order-in-time  Schr\"{o}dinger-like equation. In particular, we saw in Figs.~\ref{fig:PartialvsFirstSpeed5000} and \ref{fig:PartialvsFirstReflectivity98} that the approximation became better when the membrane reflectivity and speed are low. However, apart form dropping higher derivatives, it is not clear where in the derivation of the first order equation Eq. (\ref{eq:PartialFirstOrder}) the restriction to small reflectivities or speeds came in. Let us develop a criterion that allows us to evaluate when the first order approximation is valid depending upon the mirror reflectivity and speed.

\begin{figure}
\includegraphics[width=\linewidth]{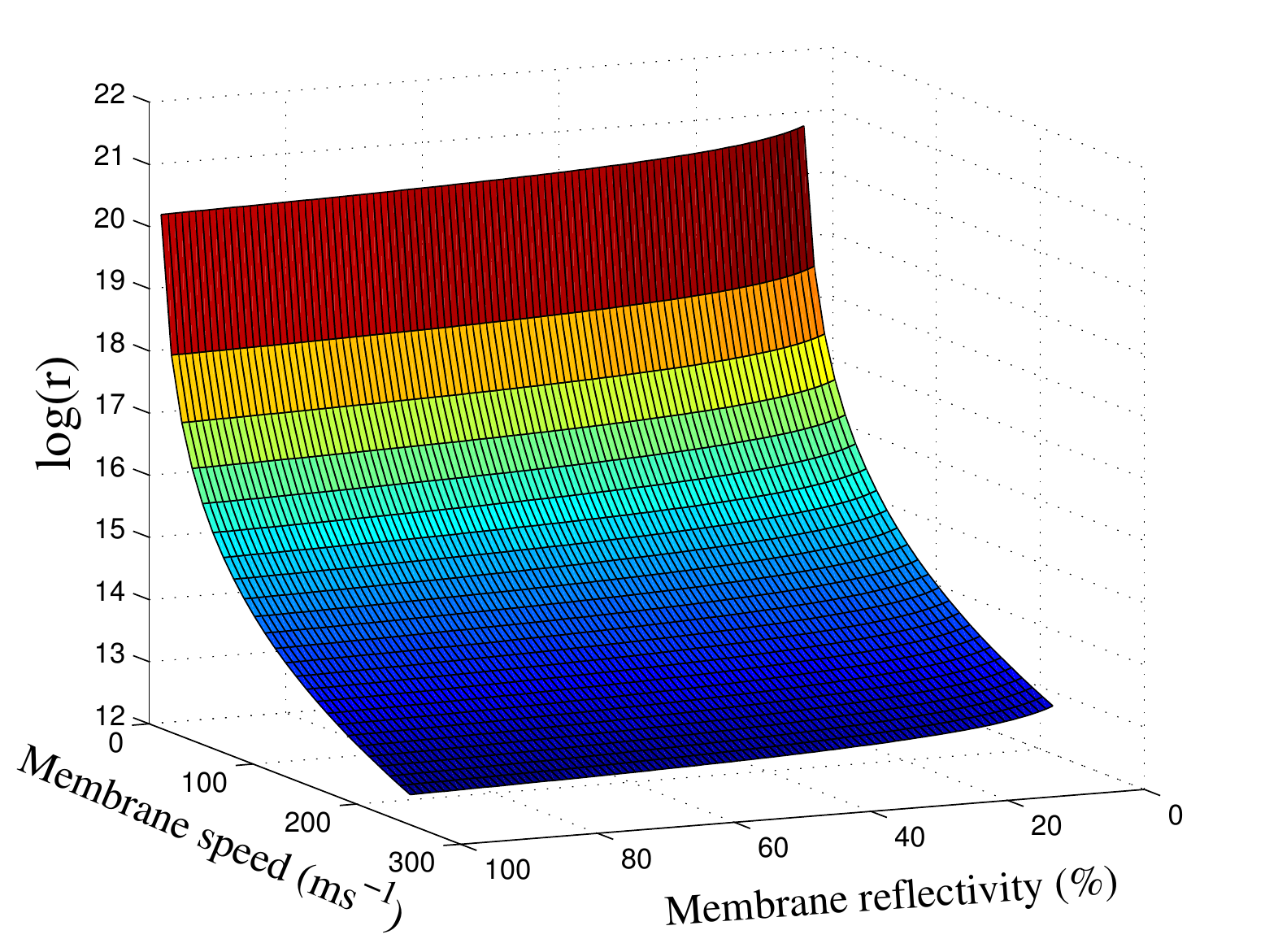}
\caption{A plot of the analytical condition given in Eq.\ (\ref{eq:ratio}) for the validity of the DFOE as a function of membrane reflectivity and speed. When $r$ is large the DFOE is a good approximation. It can be clearly seen that r becomes large at small speeds. The dependence on reflectivity is much gentler, but there is still a discernible monotonic increase in $r$ as the reflectivity is reduced. These trends are in agreement with the numerical results shown in previous sections, showing that the DFOE becomes a better approximation at low membrane speed and reflectivity.} \label{eq:ratio condition}
\end{figure}

Comparing Eqns.\ (\ref{eq:PartialSecondOrder}) and (\ref{eq:DSOEFactorize}), we see that the first order approximation is equivalent to solving the equation 
\begin{equation}
\frac{d^2}{dt^2}  
 \left(
\begin{array}{c}a_{L}\\
a_{R}
\end{array}  \right) 
 = \left( - H_{\mathrm{LZ}}^2 + i  \dot{H}_{\mathrm{LZ}}     \right)  \left(
\begin{array}{c}a_{L}\\
a_{R}
\end{array}  \right) 
\label{eq:DerivativeOfFirsOrder}
\end{equation}
which differs from the true equation by the term 
\begin{equation}
 i  \dot{H}_{\mathrm{LZ}}  = i\left(
 \begin{matrix}
\dot{\Gamma} & 0 \\
0 & -\dot{\Gamma}
\end{matrix}
\right). 
\end{equation}
Thus, for the DFOE to be a valid approximation to the DSOE, we require that the ratio $r \equiv ||H_{\mathrm{LZ}}^2||^2/||\dot{H}_{\mathrm{LZ}}||^2$ be large, i.e.\  $||H_{\mathrm{LZ}}^2||$ be much larger than  $||\dot{H}_{\mathrm{LZ}}||$.
The symbol $||\cdot||$ represents the norm of the matrix given by the square root of the  sum of each matrix element squared \cite{SpenceFriedberg}. Substituting in $H_{\mathrm{LZ}}$ and $\dot{H}_{\mathrm{LZ}}$, the ratio is given by
\begin{eqnarray}
r  = & &\frac{8\Delta^2\omega_{\mathrm{av}}^2+(\left[\omega_{\mathrm{av}}+\Gamma\right]^2+\Delta^2)^2+(\left[\omega_{\mathrm{av}}-\Gamma\right]^2+\Delta^2)^2}{2\dot{\Gamma}^2} \\
=& &  \frac{(\gamma^{2}\Delta L^{4}+\Delta^{4}+\omega_{\mathrm{av}}^{4})+6\omega_{\mathrm{av}}^{2}(\Delta^{2}+\gamma\Delta L^{2})+2\gamma\Delta^{2}\Delta L^{2}}{\gamma\dot{\Delta L}^{2}}  \label{eq:ratio}
\end{eqnarray}
where the second line follows from putting $\Gamma=\sqrt{\gamma} \, \Delta L$ and simplifying. 

The role of the optical frequency and mirror speed in the validity of the DFOE is quite clear from this expression for $r$: increasing $\omega_{\mathrm{av}}$ and decreasing $\dot{\Delta L}$ contribute to increasing $r$. What is not as obvious is the role of the reflectivity which according to Eqs.\ (\ref{eq:deltadef}) and (\ref{eq:gammadef})  appears in the terms  $\Delta$ and $\gamma$ through their dependence upon $\alpha$. Intuitively, it seems that a higher reflectivity causes the membrane to perturb the field more and should therefore should lead to a breakdown of the first order approximation. That this is indeed the case can be demonstrated by differentiating $r$ with respect to $\alpha$, from which we find that the derivative is always negative showing that $r$ monotonically decreases as $\alpha$ (and hence $R$) increases. The dependence of $r$ on reflectivity and speed are shown in Fig.\ \ref{eq:ratio condition}.

A further pictorial explanation can be found in the structure of the frequencies $\omega_{2/1}$ near an avoided crossing as shown in Fig.~\ref{fig:AvoidedCrossing}. One can see that as the central membrane reflectivity approaches unity the avoided crossing curves become steeper (asymptotically approaching the diabatic frequencies given by the red dashed curves) and change very rapidly at the avoided crossing itself as $\Delta$ shrinks. This implies a faster change of the frequencies [and quantities such as $\beta$ given in Eq.\ (\ref{eq:betadef})] with membrane position and hence that second order derivatives become more important in this limit.

\section{Radiation Pressure and the Work-Energy Theorem}
\label{sec:RadiationPressure}

In this section we attempt to give a more physical explanation for the change in energy of the electromagnetic field seen in the second-order-in-time descriptions of the dynamics (ASOE and DSOE). By applying the work-energy theorem $\Delta \mathcal{E} = \mathcal{W} = \int \mathbf{F} \cdot \mathrm{d}\mathbf{x}$, we show that the radiation pressure exerted by the field on the membrane fully accounts for the changes in electromagnetic energy we have computed in Sections \ref{sec:adiabatictheorem} and \ref{sec:LocalModeDynamics}. This also provides a self consistency check on our numerical simulations. Starting from the Maxwell stress tensor, we carefully derive the radiation pressure of light in the two mode approximation near an avoided crossing. We show that the radiation pressure obtained by simply adding the pressures due to each adiabatic mode [$U_{1/2}(x,\Delta L)$] individually leads to erroneous results and is not equivalent to the radiation pressure applied by the net electric field that includes interference. 

\begin{figure}[tp]
\includegraphics[width=\linewidth]{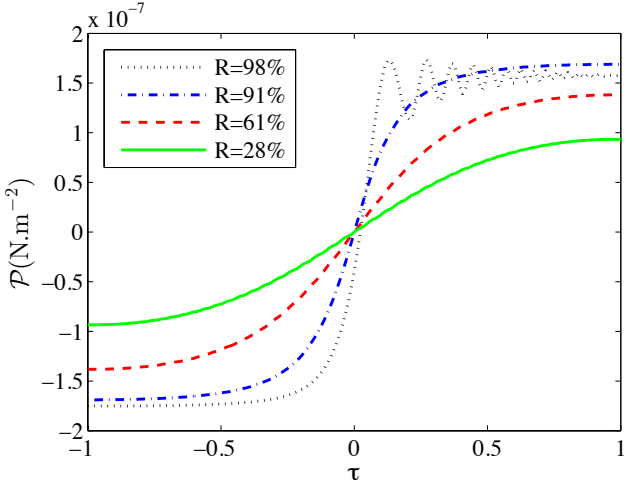}
\caption{Evolution of the radiation pressure on the central membrane during passage through an avoided crossing for four different membrane  reflectivities. The membrane speed is held constant at 5000 ms$^{-1}$ and the radiation pressure is calculated using (\ref{eq:WorkII}).  The maximum radiation pressure is greater at larger membrane reflectivities, however, at $98 \%$ reflectivity the radiation pressure exhibits oscillatory behaviour due to the non-adiabatic nature of the optical dynamics.}
\label{fig:RadPress1}
\end{figure}

The effect of radiation pressure can in fact be seen in  Figs.~\ref{fig:PartialvsFirstSpeed5000} and \ref{fig:PartialvsFirstReflectivity98} where the light is initially localized on the right side of the cavity and the membrane is moved from left to right at a constant speed. The radiation pressure pushes against the membrane, and hence, to maintain a constant speed, we need to apply a force equal in magnitude to the radiation force, but in the opposite direction. Therefore, positive work is done by the membrane  on the optical field and the latter's energy will increase. Furthermore, one can see in Figs.~\ref{fig:PartialvsFirstSpeed5000} and \ref{fig:PartialvsFirstReflectivity98} that the energy pumped in reaches a maximum value. This occurs at the point where the light intensities on the left and right sides of the membrane are equal and the radiation pressure cancels, as can be seen in Fig.~\ref{fig:RadPress1} which plots the radiation pressure corresponding to the various curves in Fig.~\ref{fig:PartialvsFirstSpeed5000}. Past this point the light intensity is greater on the left and the radiation pressure points in the same direction as the membrane motion which means that light does work upon the membrane. An external force has to be applied in the opposite direction to the membrane motion in order to maintain a constant speed. 

The force due to radiation pressure on some region of volume $\mathcal{V}$ and surface area $\mathcal{S}$ is given by \cite{Griffiths}
\begin{equation}
\mathbf{F}=\int_{\mathcal{S}}\overleftrightarrow{\mathbf{T}}\cdot \mathrm{d}\mathbf{a}-\frac{\partial}{\partial t}\int_{\mathcal{V}}\epsilon(\mathbf{r})\mathbf{E}\times\mathbf{B} \, 
\mathrm{dV} \label{eq:radiationpressure}
\end{equation}
where $\overleftrightarrow{\mathbf{T}}$ is the Maxwell stress tensor defined by
\begin{equation}
T_{ij}\equiv\epsilon_{0} \left(E_{i}E_{j}-\frac{1}{2}\delta_{ij}E^{2}\right)+\frac{1}{\mu_{0}}\left(B_{i}B_{j}-\frac{1}{2}\delta_{ij}B^{2}\right).
\end{equation}
We note in passing that the second term on the right hand side of Eq.\ (\ref{eq:radiationpressure}) is responsible for the difference between the Abraham and Minkowski expressions for the momentum of light in a medium \cite{Hinds05}. We shall neglect it here because in the $\delta$-function membrane model the volume $\mathcal{V}$ is vanishingly small. The first term, on the other hand, depends on the surface $\mathcal{S}$ of the membrane interfaces and this does not vanish. Since the only non-zero components of the electromagnetic field are $E_z$  and $B_{y}$, the stress tensor is purely diagonal. Furthermore, we only require the force along the $x$-axis and thus the only component of $\overleftrightarrow{T}$ we need is $T_{xx}$ which is given by
\begin{equation}
T_{xx}=-\frac{\epsilon_{0}}{2}E_{z}^{2}-\frac{1}{2\mu_{0}}B_{y}^{2}.
\end{equation}
Hence, the force on the membrane is 
\begin{equation}
F=\int_{\mathcal{S}}T_{xx}\mathrm{d}a_{x}=\mathcal{A}\left\{ -\frac{\epsilon_{0}}{2}E_{z}^{2}-\frac{1}{2\mu_{0}}B_{y}^{2}\right\} _{\mathrm{left}}^{\mathrm{right}}
\end{equation}
where $\mathcal{A}$ is the transverse area of the cavity mode at the membrane and the limits are evaluated at the left and right interfaces of the membrane. It is useful to first picture this for the case of a membrane of finite width and then take the limit as the width shrinks to zero. The radiation pressure is therefore
\begin{align}
\mathcal{P}=\frac{F}{\mathcal{A}} & =\left\{ -\frac{\epsilon_{0}}{2}E_{z}^{2}-\frac{1}{2\mu_{0}}B_{y}^{2}\right\} _{\mathrm{left}}^{\mathrm{right}} \nonumber \\
& =-\frac{1}{2\mu_{0}}\left\{ B_{y}^{2}\right\} _{\mathrm{left}}^{\mathrm{right}}
\end{align}
and is simply proportional to the difference of the magnetic field intensity between the two sides. The electric field does not contribute because it is continuous at the membrane interface.  By contrast, the magnetic field is discontinuous because it is related to the spatial derivative of the electric field, and for a $\delta$-function dielectric $\partial_{x} E$ has finite jump across it. 

With this expression for the radiation pressure in hand, the work-energy theorem predicts that the change in the energy of the electromagnetic field will be
\begin{equation}
\frac{\Delta \mathcal{E}(\tau)}{\mathcal{A}}=-v\int_{-1}^{\tau} \mathcal{P}(\tau^{\prime}) \, \mathrm{d}\tau^{\prime} 
\label{eq:WorkI}
\end{equation}
where $\tau$ is defined in Sec.~\ref{sec:adiabatictheorem} and the negative sign recognizes the fact that we need the work done on the field by the membrane rather than vice versa. Once the magnetic field has been computed, the radiation pressure interpretation of the physical mechanism behind the energy change can be verified by comparing it against Eq.~(\ref{eq:lightenergy}) which gives
\begin{equation}
\frac{\Delta \mathcal{E}(\tau)}{\mathcal{A}}=\epsilon_{0}\left\{ |a_{1}(\tau)|^{2}+|a_{2}(\tau)|^{2}-1\right\}
\label{eq:energyformula}
\end{equation}
assuming that we pick the initial amplitude sum to be $1$.

\begin{figure}[tp]
\includegraphics[width=\linewidth]{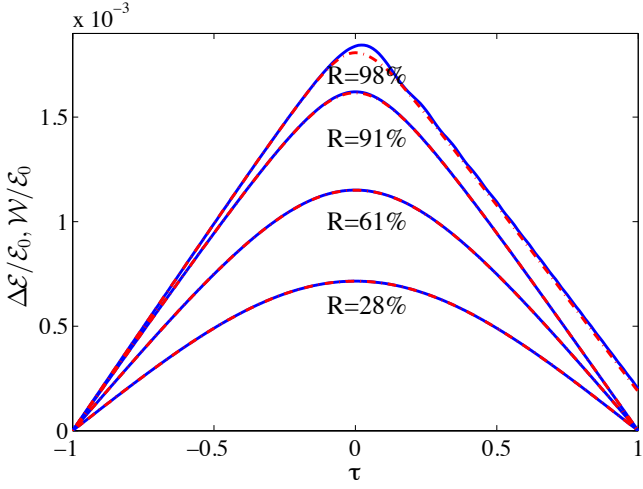}
\caption{Comparison of the change in energy $\Delta \mathcal{E}$ of the optical field (solid blue curves) with the work done $\mathcal{W}$ by radiation pressure on the membrane (red dash-dot curves) during passage through an avoided crossing. Both quantities are in units of the initial energy $\mathcal{E}_{0}$. Eq.\ (\ref{eq:energyformula}) is used to calculate  $\Delta \mathcal{E}$ and the radiation pressure is obtained by summing the contributions given by Eq.\ (\ref{eq:pressuremodem}) for the two modes separately.   The agreement is good but breaks down at higher membrane reflectivities. The membrane speed is 5000 ms$^{-1}$. The same mode amplitudes $\{c_{1}(t),c_{2}(t)\}$ were used for both sets of curves and were calculated using the ASOE. As usual, the light is initially localized on the right hand side of the membrane. }
\label{fig:WorkEnergyII}
\end{figure}

\begin{figure}[tp]
\includegraphics[width=\linewidth]{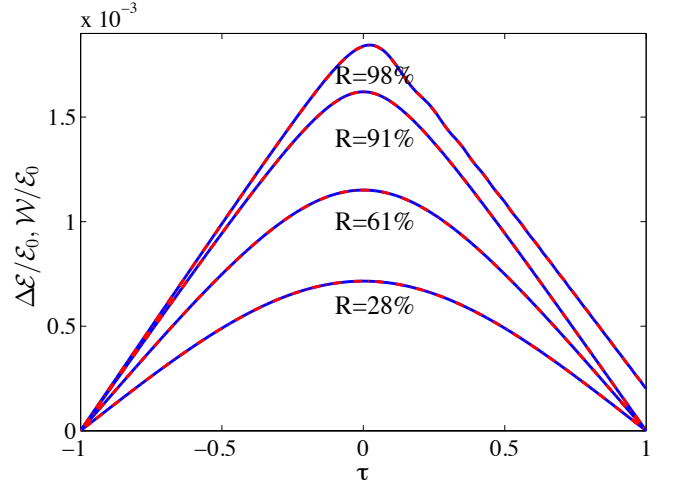}
\caption{Same as Fig.\ \ref{fig:WorkEnergyII}, except that the radiation pressure is calculated using Eq.\ (\ref{eq:WorkII}) which includes interference between the contributions from each mode. As can be seen, the agreement is excellent at all reflectivities.}
\label{fig:WorkEnergyI}
\end{figure}

The magnetic field entering the expression for radiation pressure can be obtained from the electric field using Maxwell's equations. The electric field  due to the $m^{\mathrm{th}}$ mode is
\begin{equation}
\mathbf{E}_{m}(x,t)=c_{m}(t) U_{m}(x,t) \exp[-i\theta_{m}(t)] \, \hat{\mathbf{z}} 
\end{equation}
which gives rise to the displacement field $\mathbf{D}(x,t)= \epsilon(x,t) \mathbf{E}(x,t)$. According to the Maxwell equation
\[
\nabla\times\mathbf{H}=\frac{\di\mathbf{D}}{\di t}
\]
the magnetizing field satisfies 
\begin{multline}
\di_{x}H_{m}(x,t)\hat{\mathbf{z}} =  \bigg\{-i \omega_{m}(t) \epsilon(x,t) c_{m}(t) U_{m}(x,t)  \\  +  \frac{\di}{\di t}[\epsilon(x,t) c_{m}(t)U_{m}(x,t)]\bigg\}  \exp[-i\theta_{m}(t)] \hat{\mathbf{z}} . \label{eq:Hfieldderivative}
\end{multline}
Given the large magnitude of the optical frequency $\omega_{m}$, we can to a very good approximation ignore the second term on the right hand side. Incorporating the boundary conditions at the end mirrors, the solutions to Eq.\ (\ref{eq:Hfieldderivative}) can be written
\[
H_{m}(x,t)=i c \epsilon_0 c_{m}(t) \exp[-i\theta_{m}(t)]G_{m}(x, \Delta L) 
\]
where
\begin{equation}
G_{m}(x,\Delta L)=
\begin{cases} 
A_{m}\cos[k_{m}(x+L_{1})], & -L_{1} \le x \le 0\\
B_{m}\cos[k_{m}(x-L_{2})], & 0 \le x \le L_{2}
\end{cases}
\end{equation}
[compare with \ Eq.\ (\ref{eq:globalmodes}), in particular, the  amplitudes $A_{m}$ and $B_{m}$ are the same as for the electric field modes $U_{m}(x,\Delta L)$].
The radiation pressure on the membrane located at $x=0$ due to a monochromatic field of frequency $\omega_{m}$  is, therefore,
\begin{align}
\mathcal{P}_{m} &= -\frac{\mu_{0}}{2}\left\{ \vert H_{m} \vert^{2}\right\} _\mathrm{left}^\mathrm{right} \nonumber \\
&=-\frac{\epsilon_{0}}{2} \vert c_{m}(t) \vert^2 \left\{B_{m}^{2}\cos^{2}(k_{m}L_{2}) - A_{m}^{2}\cos^{2}(k_{m}L_{1}) \right\} .
\label{eq:pressuremodem}
\end{align}

\begin{figure}[tp]
\includegraphics[width=\linewidth]{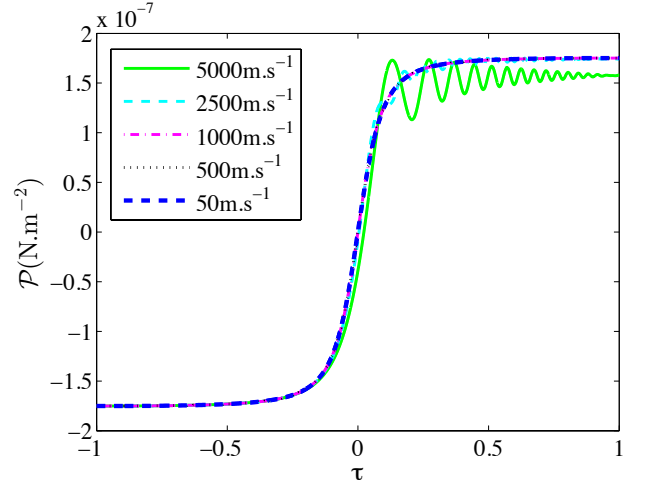}
\caption{Same as Fig.\ \ref{fig:RadPress1} except that here we fix the membrane reflectivity at $98 \%$ but choose five different membrane speeds. The highest mirror speed leads to non-adiabatic optical dynamics and consequently an oscillatory behaviour of the radiation pressure. This figure corresponds to the optical dynamics shown in Figs.\ \ref{fig:AdiabaticCriteriaSingleMode2} and \ref{fig:AdiabaticCriteriaSingleMode}.}
\label{fig:RadPress2}
\end{figure}

In Fig.~\ref{fig:WorkEnergyII} we compare the results of the radiation pressure calculation with the exact result given in Eq.\ (\ref{eq:energyformula}). The total radiation pressure is taken to be the sum of that due to each  monochromatic light field separately. We see that the agreement is excellent for the lower reflectivity cases, but there are noticeable differences between the $98 \%$ reflectivity curves. This is because we are not including interference between the two modes involved in the avoided crossing. Rather of summing up the forces due to individual frequencies of light, let us instead find the force due to the net electromagnetic field. The total electric field is
\begin{multline}
\mathbf{E}(x,t)=  \left\{ c_{1}(t)\exp[-i\theta_{1}(t)]U_{1}(x,t)+ \right.  \\ \left. c_{2}(t)\exp[-i\theta_{2}(t)]U_{2}(x,t) \right\} \hat{\mathbf{z}}.
\end{multline}
Following the analogous steps as the single mode case, we find that the magnetizing field due to two modes is
\begin{multline}
H(x,t) = i c \epsilon_0 \left\{ c_{1}(t)\exp[-i\theta_{1}(t)]G_{1}(x,\Delta L)+ \right. \\ \left. c_{2}(t)\exp[-i\theta_{2}(t)]G_{2}(x, \Delta L)\right\}
\end{multline}
and therefore the radiation pressure is given by
\begin{multline}
\mathcal{P} =-\bigg\{ \vert c_{1}(t) \vert^2 \left( B_{1}^{2}(t) \cos^{2}[k_{1} L_{2} ]-A_{1}^{2}(t) \cos^{2}[k_{1} L_{1} ] \right) \\
+\vert c_{2}(t) \vert^2 \left( B_{2}^{2}(t) \cos^{2}[k_{2} L_{2} ]-A_{2}^{2}(t) \cos^{2}[k_{2} L_{1} ] \right) \\
+2 \Re \left[  c_{1}^{\ast}(t) c_{2}(t) e^{i \theta_{12}} \right] \left( B_{1}(t)B_{2}(t) \cos [k_{1} L_{2}]\cos[k_{2} L_{2} ] \right. \\
- \left. A_{1}(t)A_{2}(t) \cos [k_{1} L_{1}]\cos[k_{2} L_{1} ] \right) \bigg\} \frac{\epsilon_{0}}{2} .
\label{eq:WorkII}
\end{multline}
The cross terms on the third and fourth lines are not included in Fig.\ \ref{fig:WorkEnergyII}, but are included in Fig.~\ref{fig:WorkEnergyI} where we find perfect agreement with the general result given in Eq.\ (\ref{eq:energyformula}).

It is instructive to plot the radiation pressure itself during passage through an avoided crossing, and this is done in Figs.\ \ref{fig:RadPress1} and \ref{fig:RadPress2} where we now exclusively use the more accurate form for the radiation pressure given in Eq.\ (\ref{eq:WorkII}). Initially the light is localized on the right of the membrane producing a radiation pressure in the $-x$ direction; on the other side of the avoided crossing the light has swapped sides and so the radiation pressure reverses direction. In Fig.\  \ref{fig:RadPress1} the effect of changing the reflectivity of the membrane is shown. As expected, the maximum radiation pressure increases with reflectivity and thus it is possible to do more work on the optical field in the regime of high reflectivity. In Fig.\  \ref{fig:RadPress2} we see the effect of varying the membrane speed. The pressure curves do not all pass through zero at the same point, there being a slight lag at higher speeds.  Perhaps the most striking feature of both Figs.\ \ref{fig:RadPress1} and \ref{fig:RadPress2} is that at higher reflectivities and speeds the radiation pressure develops oscillations.  If the transfer is adiabatic then light is smoothly transferred from one side of the cavity to the other with the radiation pressure monotonically reversing direction. However, non-adiabatic passage means that not all the light is transferred  to the other side. Instead, the system is left in an ``excited'' state with a certain fraction of the light sloshing back and forth between the two sides of the cavity leading to an oscillatory radiation pressure.

\section{Quantization}
  \label{sec:DCE}

Although the focus of this paper is on classical fields, in this section we review the quantum version of the problem  in order to better understand the connection between the two. In Section \ref{sec:setup} we discussed quantization for the case of a static membrane and showed how the second-order-in-time wave equation became a first order Heisenberg equation of motion for the field operator $\hat{C}(t)$. The dynamic membrane case is more involved and our approach here, which makes use of Dirac's canonical quantization method, is adapted from Law's treatment of a single cavity with a moving end mirror \cite{Law1} (in reference \cite{Law3} Cheung and Law treat the problem of the double cavity but they consider the membrane position and momentum as dynamical variables to be included in the Hamiltonian rather than following a prescribed motion as we do here). As for the static case, the quantum operators in the Heisenberg representation can be constructed from the solutions to the classical wave equation. It is more usual to work with the vector potential $A(x,t)$ than the electric field; the former satisfies the wave equation 
\begin{equation}
\frac{\partial^2 A}{\partial x^2}-\frac{\partial}{\partial t} \left[ \mu_{0}  \epsilon(x,t) \frac{\partial A}{\partial t} \right]=0 
\label{eq:waveeqnforA}
\end{equation}
with the same boundary conditions at the end mirrors as the electric field. Unlike in the static case, it is necessary to introduce an auxiliary  variable $\pi(x,t) \equiv \epsilon(x,t) \partial A(x,t)/ \partial t$ which is the `momentum' conjugate to $A(x,t)$. Canonical quantization is achieved by imposing the commutation relation $[ \hat{A}(x,t), \hat{\pi}(x',t)]=i \hbar \delta(x-x')/\mathcal{A}$, where $\mathcal{A}$ is the area of the mode functions. 

A separation of variables can be achieved by expanding $\hat{A}(x,t)$ and  $\hat{\pi}(x,t)$ over the adiabatic modes with time-dependent amplitudes $\hat{Q}_{n}(t)$ and $\hat{P}_{n}(t)$: 
\begin{eqnarray}
\hat{A}(x,t) & = & \frac{1}{\sqrt{\epsilon_{0}}} \sum_{n} \hat{Q}_{n} (t) U_n(x,t) \label{eq:Aexpansion} \\
\hat{\pi}(x,t) & = & \frac{\epsilon(x,t)}{\sqrt{\epsilon_{0}}} \sum_{n} \hat{P}_{n} (t) U_n(x,t).  \label{eq:piexpansion}
\end{eqnarray}
$\hat{Q}_{n}(t)$ and $\hat{P}_{n}(t)$ play roles analogous to the canonical position and momentum variables of the harmonic oscillator and indeed they obey the canonical commutator $[\hat{Q}_{m}(t),\hat{P}_{n}(t)]=i (\hbar/\mathcal{A}) \delta_{m,n}$,  a relation which is inherited from that between $\hat{A}(x,t)$ and $\hat{\pi}(x,t)$. In the static case $\hat{P}_{n}(t)$ can be eliminated in favor of $\hat{Q}_{n}(t)$, however, the fact that the separation into space- and time-dependent variables  is not complete in the dynamic case (because the mode functions also depend on time), introduces an extra coupling between $\hat{Q}_{n}(t)$ and $\hat{P}_{n}(t)$ that prevents a description purely in terms of  $\hat{Q}_{n}(t)$ and hence in terms of a single first-order-in-time equation for $\hat{Q}_{n}(t)$. Explicit expressions for $\hat{Q}_{n}(t)$ and $\hat{P}_{n}(t)$ can be found by inverting the above equations by using the orthonormality of the mode functions [Eq.\ (\ref{eq:orthonormal})] giving
\begin{eqnarray}
\hat{Q}_{n}(t) & = & \frac{1}{\sqrt{\epsilon_{0}}} \int_{-L_{1}}^{L_{2}} \epsilon(x,t) \hat{A}(x,t) U_{n}(x,t) \ d x \label{eq:Qn} \\
\hat{P}_{n}(t) & = & \frac{1}{\sqrt{\epsilon_{0}}} \int_{-L_{1}}^{L_{2}}  \hat{\pi}(x,t) U_{n}(x,t) \ d x. \label{eq:Pn}
\end{eqnarray}
Equations of motion for $\hat{Q}_{n}(t)$ and $\hat{P}_{n}(t)$ are obtained by taking time derivatives of these expressions (details are given in Appendix \ref{app:quantumEOM})
\begin{eqnarray}
\frac{d \hat{Q}_n(t)}{dt} & = & \hat{P}_{n}(t)- \sum_{m}  G_{nm} (t) \hat{Q}_{m} (t) \label{eq:dQdt} \\
\frac{d\hat{P}_n(t)}{dt} & = & - \omega_{n}^{2}(t) \hat{Q}_{n}(t)+ \sum_{m} G_{mn} (t) \hat{P}_{m} (t)\label{eq:dPdt}
\end{eqnarray}
where $G_{nm}(t)= \dot{q} g_{nm}(q) $ and
\begin{equation}
g_{nm}(q) = \int_{-L_{1}}^{L_{2}} \frac{\epsilon(x,q)}{\epsilon_{0}} U_{n}(x,q) \frac{\partial U_{m}(x,q)}{\partial q} dx . \label{eq:gdefinition}
\end{equation}
To keep this expression compact we have introduced the symbol $q$ for the membrane displacement $\Delta L/2$. It is clear that the membrane motion introduces coupling between $\hat{Q}_{n}(t)$ and $\hat{P}_{n}(t)$ that is absent in the static case. The coupling is governed by $G_{nm}(t)$ and is directly proportional to the velocity of the membrane $v= \dot{q}$.

By integrating the coupled equations of motion Eqns.\ (\ref{eq:dQdt}) and ($\ref{eq:dPdt}$) forward in time the quantum dynamics of the electromagnetic field can be calculated from given initial conditions. However, in order to gain physical insight it is useful to find the corresponding Hamiltonian, i.e.\ the Hamiltonian that gives $d\hat{Q}_{n}/dt$ and $d\hat{P}_{n}/dt$  as its equations of motion via the Heisenberg equation $i \hbar d\hat{O}/dt=[\hat{O},\hat{H}]$, where $\hat{O}$ stands for either $\hat{Q}$ or $\hat{P}$. One can verify that the Hamiltonian that does the trick is \cite{Law1}
\begin{eqnarray}
\hat{H} & = &\frac{1}{2}\sum_{n} \bigg[ \hat{P}_{n}^2+\omega_{n}^{2}(t)\hat{Q}_{n}^2 
-G_{nn}(t) \left(\hat{P}_{n} \hat{Q}_{n}+\hat{Q}_{n}\hat{P}_{n}\right) \bigg] \nonumber \\
&& -\sum_{m \neq n} G_{m,n}(t) \hat{P}_{m} \hat{Q}_{n} \ . \label{eq:QuantumHamiltonian1}
\end{eqnarray}
The first two terms describe a harmonic oscillator with a parametrically driven frequency. The third term introduces correlations in phase space that produce a ``squeezing effect'' \cite{schutzhold98} and the final term introduces further correlations that have been called the ``acceleration effect'' \cite{schutzhold98}, even though a constant velocity is enough (there is no acceleration in the particular cases we have considered in the earlier sections of this paper). The squeezing and acceleration effects give rise to field dynamics such as parametric amplification and transfer of excitations between modes. Indeed, the squeezing term in the Hamiltonian corresponds to that of a degenerate parametric amplifier \cite{gardiner}. 

Parametric amplification/reduction is most clearly seen in the Hamiltonian if it is expressed
in terms of the annihilation and creation operators defined as
 \begin{eqnarray}
\hat{C}_{n}(t) \equiv \frac{1}{\sqrt{2 \hbar \omega_{n}(t)}}\left( \omega_n(t)\hat{Q}_n(t)+i \hat{P}_{n}(t) \right) \label{eq:annihilationdef} \\
\hat{C}^{\dag}_{n}(t) \equiv \frac{1}{\sqrt{2 \hbar \omega_{n}(t)}}\left( \omega_n(t)\hat{Q}_n(t)-i \hat{P}_{n}(t) \right)  \label{eq:creationdef}
\end{eqnarray}
which annihilate and create photons in the $n^{\mathrm{th}}$ adiabatic mode.
To find the Hamiltonian we proceed similarly to before by taking time derivatives of the expressions for $\hat{C}_{n}$ and $\hat{C}^{\dag}_{n}$ to obtain their equations of motion in terms of $d\hat{Q}_{n}/dt$ and $d\hat{P}_{n}/dt$ whose expressions are already known  and then inferring the Hamiltonian that generates them. The resulting Hamiltonian is \cite{Law1}
\begin{equation}
\hat{H}  =  \sum_n \hbar \omega_n(t) \hat{C}_{n}^{\dag} \hat{C}_{n} -\frac{i \hbar \dot{q}}{2} \hat{\Lambda}(q)  \label{eq:QuantumHamiltonian2}
\end{equation}
where
\begin{eqnarray}
\hat{\Lambda}(q) & = &
 \sum_{n} \left\{ g_{nn}(q)-\frac{1}{2 \omega_{n}} \frac{\partial \omega_{n}}{\partial q}\right\} \left[ \left(\hat{C}^{\dag}_{n}\right)^2 - \hat{C}_{n}^{2} \right] \label{eq:Lambda1}  \\
&&  +\sum_{m \neq n} \sqrt{\frac{\omega_{m}(q)}{\omega_{n}(q)}} g_{mn}(q) \left( \hat{C}^{\dag}_{m}   \hat{C}^{\dag}_{n} + \hat{C}^{\dag}_{m}   \hat{C}_{n} - \mathrm{h.c.}   \right) \nonumber
\end{eqnarray}
in which h.c. stands for hermitian conjugate. $ \hat{\Lambda}(q)$ contains the terms responsible for non-adiabatic transfer (scattering of photons between modes) and also amplification/reduction processes. Due to its prefactor of $\dot{q}$, these terms arise purely as a result of membrane motion. The diagonal terms in $\hat{\Lambda}(t)$ give rise to single-mode squeezing by creating and annihilating photons in pairs in the same mode, whereas the off-diagonal ($m \neq n$) terms give rise both to two-mode squeezing ($  \hat{C}^{\dag}_{m}   \hat{C}^{\dag}_{n} -$ h.c. terms) where pairs of photons are created and annihilated in different modes, and scattering between modes ($ \hat{C}^{\dag}_{m}   \hat{C}_{n} - $ h.c. terms). We saw the classical analogues of these processes in Sections \ref{sec:adiabatictheorem}, \ref{sec:LocalModeDynamics} and \ref{sec:RadiationPressure} where we found both the transfer of field energy between modes and the amplification/reduction of the total energy in both modes even in the absence of transfer. However, unlike in the classical case, in quantum mechanics photons can be created from the vacuum, and this is the DCE.  

In the two-mode case the Hamiltonian can be written out explicitly. In order to obtain analytic expressions for the coefficients one can approximate the adiabatic mode functions by superpositions of modes perfectly localized in either the left or right sides of the cavity, as detailed in Appendix \ref{app:analyticexpressions}. One finds
\begin{eqnarray}
g_{11} & = & g_{22}=0 \\
g_{12} & = & -g_{21}= -\frac{d \Gamma (q)}{dq} \frac{\Gamma(q) \Delta}{2 (\Gamma^2(q)+\Delta^2)^{3/2}}
\end{eqnarray}
where $\Gamma(q)= 2\sqrt{\gamma}q \approx 2 (\omega_{\mathrm{av}}/L) q$ so that $d \Gamma/ dq \approx 2 \omega_{\mathrm{av}}/L$. In addition, at this level of approximation 
\begin{equation}
\frac{1}{\omega_{1}}\frac{d \omega_{1}}{ dq} \approx -\frac{1}{\omega_{2}}\frac{d \omega_{2}}{ dq} = \frac{4 \omega_{\mathrm{av}}}{L \sqrt{\Delta^2+\Gamma^2(q)}}\frac{q}{L}  
\end{equation}
and 
\begin{eqnarray}
\sqrt{\frac{\omega_{1}(q)}{\omega_{2}(q)}} \approx 1- \frac{\sqrt{\Delta^2+\Gamma^2(q)}}{\omega_{\mathrm{av}}} \\
\sqrt{\frac{\omega_{2}(q)}{\omega_{1}(q)}} \approx 1+ \frac{\sqrt{\Delta^2+\Gamma^2(q)}}{\omega_{\mathrm{av}}}.
\end{eqnarray}
As shown in Appendix \ref{app:analyticexpressions}, the corrections to unity in these latter two expressions are necessary to consistently keep terms of the same magnitude as $(1/\omega_{1}) d \omega_{1}/dq$. The two-mode quantum Hamiltonian in the adiabatic basis then takes the form
\begin{equation}
\hat{H}=\hbar \omega_{1}(t) \hat{C}_{1}^{\dag}\hat{C}_{1} +\hbar \omega_{2}(t) \hat{C}_{2}^{\dag}\hat{C}_{2}  -\frac{i \hbar \dot{q}}{2} \hat{\Lambda}(q)  \label{eq:QuantumHamiltonian3}
\end{equation}
where
\begin{eqnarray}
\hat{\Lambda}(q) & = & \frac{1}{2 \omega_{1}}\frac{d \omega_{1}}{dq} \left( \hat{C}_{2}^{\dag} \hat{C}_{2}^{\dag} - \hat{C}_{2} \hat{C}_{2} +  \hat{C}_{1} \hat{C}_{1}  - \hat{C}_{1}^{\dag} \hat{C}_{1}^{\dag}  \right) \nonumber \\
& & + 2 g_{21}\bigg\{ \left( \hat{C}_{2}^{\dag} \hat{C}_{1} - \hat{C}_{1}^{\dag} \hat{C}_{2} \right) \nonumber \\
&& \quad +\frac{\sqrt{\Delta^2+\Gamma^2(q)}}{\omega_{\mathrm{av}}} \left(\hat{C}_{2}^{\dag}\hat{C}_{1}^{\dag}- \hat{C}_{2}\hat{C}_{1} \right)\bigg\}.
 \label{eq:Lambda3}
\end{eqnarray}
We note that the squeezing terms [that appear on the first and third lines of $\hat{\Lambda}(q)$] are weaker by a factor of $\sim \Delta/\omega_{\mathrm{av}}$ than the intermode transfer terms [that appear on the second line of $\hat{\Lambda}(q)$]. 

Finally, in order to compare the quantum field Hamiltonian with that of Landau-Zener problem, let us re-write it in the diabatic basis. This can be done by rotating the operators  (in the Schr\"{o}dinger representation) as
\begin{eqnarray}
\hat{C}_{1} & = & \sin \theta \ \hat{a}_{R}+ \cos \theta \ \hat{a}_{L} \\
\hat{C}_{2} & = & \cos \theta \ \hat{a}_{R} - \sin \theta \ \hat{a}_{L}
\end{eqnarray}
where $\sin \theta$ and $\cos \theta$ are defined in Eqns. (\ref{eq:sindefn}) and (\ref{eq:cosdefn}).
Making use of the following exact results:
\begin{eqnarray}
&& \cos^2 \theta - \sin^{2} \theta  =  \frac{\Gamma(q)}{\sqrt{\Delta^2+\Gamma^2(q)}} \\
&& \cos \theta \sin \theta = \frac{\Delta}{2\sqrt{\Delta^2+\Gamma^2(q)}} \\
&&  \cos \theta \sin \theta \left( \omega_{1}-\omega_{2}\right)= \Delta \\
&& \omega_{2} \cos^2 \theta + \omega_{1} \sin^{2} \theta = \omega_{\mathrm{av}} + \Gamma(q) \\
&& \omega_{2} \sin^2 \theta + \omega_{1} \cos^{2} \theta = \omega_{\mathrm{av}} - \Gamma(q) \ ,
\end{eqnarray}
we obtain the Hamiltonian
\begin{eqnarray}
\hat{H} & = & \hbar \left\{\omega_{\mathrm{av}} +\Gamma(q)\right\} \hat{a}_{R}^{\dag}\hat{a}_{R} +\hbar\left\{ \omega_{\mathrm{av}}-\Gamma(q)\right\} \hat{a}_{L}^{\dag}\hat{a}_{L} \nonumber \\
&& +\hbar \Delta (\hat{a}^{\dag}_{R} \hat{a}_{L}+\hat{a}^{\dag}_{L}\hat{a}_{R})  -\frac{i \hbar \dot{q}}{2} \hat{\Lambda}(q)  \label{eq:QuantumHamiltonian4}
\end{eqnarray}
where this time
\begin{eqnarray}
&& \hat{\Lambda}(q)  =  2 g_{21} \left( \hat{a}_{R}^{\dag} \hat{a}_{L} - \hat{a}_{L}^{\dag} \hat{a}_{R} \right) \nonumber \\ && + 
 \bigg\{ \frac{1}{2 \omega_{1}}\frac{d \omega_{1}}{dq} \frac{2 \Delta}{\sqrt{\Delta^2+\Gamma^2(q)}} -2 g_{21} \frac{\Gamma}{\omega_{\mathrm{av}}} \bigg\} \nonumber \\
 && \quad \times \left( \hat{a}_{R} \hat{a}_{L} - \hat{a}^{\dag}_{R} \hat{a}^{\dag}_{L} \right) \nonumber \\
&& + \bigg\{  \frac{1}{2 \omega_{1}}\frac{d \omega_{1}}{dq}  \frac{\Gamma(q)}{\sqrt{\Delta^2+\Gamma^2(q)}} +  g_{21} \frac{\Delta}{ \omega_{\mathrm{av}}} \bigg\} \nonumber \\
&& \quad \times \left( \hat{a}_{R}^{\dag} \hat{a}_{R}^{\dag} - \hat{a}_{R} \hat{a}_{R} +  \hat{a}_{L} \hat{a}_{L} - \hat{a}_{L}^{\dag} \hat{a}_{L}^{\dag}   \right) .
 \label{eq:Lambda4}
\end{eqnarray}
The first part of the Hamiltonian [everything except $\hat{\Lambda}(q)$] is independent of the membrane velocity and conserves total photon number. It has the structure of a many-particle version of the Landau-Zener problem: The diagonal terms feature the diabatic energies $ \hbar \{\omega_{\mathrm{av}} \pm \Gamma(q)\}$ that vary linearly with $q$, and the off-diagonal term gives the constant photon transfer rate $\Delta$ between the two diabatic modes. $\hat{\Lambda}(q)$ contains the ``beyond Landau-Zener'' effects including photon pair creation and annihilation in the form of both single and two mode squeezing, and also (photon number conserving) intermode transfer (the first line).  Current treatments of the Landau-Zener (``Photon Shuttle'') problem in the optomechanical literature \cite{HarrisPhotonShuttle} do not include pair creation and annihilation as these effects are expected to be tiny in present experimental setups; even the dominant term in $\hat{\Lambda}(q)$ is an intermode transfer term, albeit a velocity dependent one. Using the experimental numbers given in Reference \cite{Sankey2010}  we can estimate (see Appendix \ref{app:analyticexpressions}) that at membrane velocities of $10$ m/s, and at displacements of the order of half way to the next avoided crossing, this term would give a comparable contribution to that of the static membrane transfer rate $\Delta= 2 \pi \times 0.1$ MHz.

The equations of motion that arise from the two-mode Hamiltonian in the adiabatic basis given in Eqns.\ (\ref{eq:QuantumHamiltonian3}) and (\ref{eq:Lambda3}) are the quantum equivalents of our ASOE derived in Section \ref{sec:adiabaticbasisEOM}. One might guess, therefore, that the equations of motion that arise from the  two-mode Hamiltonian in the diabatic basis given in Eqns.\ (\ref{eq:QuantumHamiltonian4}) and (\ref{eq:Lambda4}) would be the quantum equivalents of the 
DSOE derived in Section \ref{sec:LocalModeDynamics}. However, this is not quite true because when deriving the DSOE we made the approximation of ignoring the time-dependence of the diabatic mode functions on the grounds that in the single particle Landau-Zener problem this is a much smaller effect than the change in amplitudes. Nevertheless, we saw in Section \ref{sec:LocalModeDynamics} that energy is not conserved by the DSOE and this can be attributed to the fact that they are second order equations in time that are not trivially first order equations that have been differentiated a second time (as shown in Section \ref{sec:ApproximationCondition}) which would conserve energy like the DFOE. Comparing with the quantum Hamiltonian in the diabatic basis, if the time-dependence of the mode functions is ignored then $g_{nm}=0$, but there is still a contribution to photon generation coming from $(1/\omega_{1}) d \omega_{1}/dq$ in $\hat{\Lambda}(q)$.

We shall not numerically solve the quantum field equations found in this section, but will leave that to a future publication. Rather, our purpose has been to understand the structure of the quantum theory in comparison to the classical one.

\section{Summary and Conclusions} 
\label{sec:conclusion}
In this paper we have examined the Landau-Zener problem in the context of an optical field whose modes undergo an avoided crossing. It can therefore be viewed as a study of adiabaticity for fields satisfying the Maxwell wave equation and is related to generalizations of the Landau-Zener theory to the many-particle case in condensed matter physics contexts \cite{Anglin03,Tomadin08,Altland08,Itin09,Oka10,Chen11,Polkovnikov11,Qian13}. By comparing the effects of successive approximations, such as ignoring the time-dependence of the modes in the diabatic basis and reducing the Maxwell wave equation to an effective Schr\"{o}dinger equation, we have emphasized some significant differences to the original Landau-Zener problem which is posed in terms of the (true) single-particle Schr\"{o}dinger wave equation. In the diabatic basis (whose modes are \emph{not} instantaneous normal modes) almost all the time-evolution occurs in the coefficients as opposed to the mode functions such that the time-evolution of the latter can be ignored. However, reducing the second order Maxwell wave equation to a first order effective Schr\"{o}dinger equation turns out to be a more severe approximation, at least conceptually, because it prevents changes in the energy of the field associated with parametric amplification (and reduction) that may be considered as classical analogues of the DCE.  The Maxwell wave equation therefore allows for a type of evolution unfamiliar from the single-particle case but which becomes particularly evident in the regime of a slowly moving membrane where the non-adiabatic transfer between the modes switches off (like in the single-particle case) and yet the total energy (i.e.\ photon population) can change. Furthermore, the energy dependence on membrane position does not vanish as the membrane velocity vanishes but tends to a fixed function that depends only on the membrane reflectivity.  This type of behaviour was explained in Section \ref{sec:RadiationPressure},  both qualitatively and quantitatively, by looking at the work done by the radiation pressure on the membrane, and this never vanishes except right at the centre of the avoided crossing. An analytic criterion [given in Eq.\ (\ref{eq:ratio})] can be derived which predicts when beyond single-particle effects become important. Apart from the expected role of the membrane velocity, i.e.\ faster membranes cause more amplification/reduction, the criterion depends on the reflectivity. A more reflective membrane perturbs the modes more, giving a sharper change in the adiabatic mode frequencies as the membrane passes through an avoided crossing.  

The criterion predicting when the single-particle picture breaks down is obtained by examining when the Maxwell wave equation can be factorized into a product of two effective Schr\"{o}dinger equations (which are Hermitian conjugates of each other). The factorization is exact for a static membrane but is  approximate in the presence of a moving membrane, as shown in Section \ref{sec:LocalModeDynamics}. This raises the question of what exactly is the connection between the effective Schr\"{o}dinger equation used to describe the classical field and the true quantum field description? The answer is rather little, at least in the moving membrane case. The effective Schr\"{o}dinger equation obtained in this paper is nothing more than an approximation to a classical field equation, and the classical field amplitude that obeys it has no interpretation in terms of a probability amplitude even though it happens to be a complex number in our treatment (the real part gives the physical electric field). Furthermore, there is only a single Schr\"{o}dinger equation for each mode [the 2x2 matrix equation given in Eq.\ (\ref{eq:PartialFirstOrder}) is for two modes].

In the true quantum field description, as given in Section \ref{sec:DCE}, each mode is described by two canonical coordinates, $\hat{Q}$ and $\hat{P}$, whose first order equations of motion [Eqns.\ (\ref{eq:dQdt}) and (\ref{eq:dPdt})] only take on the harmonic oscillator form in the limit of a stationary membrane. Only in this limit can $\hat{P}$ be eliminated to obtain the second order in time equation of motion purely in terms of $\hat{Q}$ which is that of a free harmonic oscillator. Converting the canonical coordinates to annihilation and creation operators leads to a Hamiltonian with two pieces: one piece [Eq.\ (\ref{eq:QuantumHamiltonian4})] which is straightforward generalization of the single-particle Landau-Zener hamiltonian to the many-particle case, and a second `beyond Landau-Zener' piece [Eq.\  (\ref{eq:Lambda4})] which depends linearly on the membrane velocity and includes the terms responsible for pair creation and annihilation.  The evolution of the quantum field obeys the true Schr\"{o}dinger equation
\begin{equation}
i \hbar \frac{\partial \vert \Psi (t) \rangle}{\partial t}= \hat{H}(t) \vert \Psi (t) \rangle
\label{eq:trueschrodinger}
\end{equation}
where $\hat{H}(t)$ can be any one of the Hamiltonians given in Section \ref{sec:DCE} and $\vert \Psi (t) \rangle$ is the state vector in Fock space describing the occupation of the various modes by photons. 

Coming back to the connection to Klein-Gordon equation mentioned in the Introduction, it is known that in the time-independent case it can be exactly reformulated in terms of two coupled Schr\"{o}dinger equations (see p19 of Reference \cite{Greiner}), as is to be expected in general for a second order equation. The solutions to each  Schr\"{o}dinger equation individually satisfy the Klein-Gordon equation. In the same time-independent regime the Maxwell wave equation  can be exactly reformulated in terms of a single Schr\"{o}dinger equation (for each mode)---see Sections \ref{sec:setup} and \ref{sec:LocalModeDynamics}. The difference arises because the Klein-Gordon equation describes a massive field which is in general complex whereas the Maxwell field is real: this means that the Klein-Gordon field excitations include particles and antiparticles whereas in the Maxwell case the photon is massless and is its own antiparticle. Of course, the Maxwell field can have two different polarizations (whereas the Klein-Gordon field is spinless) although we have not made use of this possibility in this work since we assumed a single linear polarization. 

A close analogy exists between the non-relativistic limit of the Klein-Gordon equation and the effective Schr\"{o}dinger equation given in Eq.\ (\ref{eq:PartialFirstOrder}) that forms the DFOE approximation used in this paper.  Substituting the ansatz $\psi(r,t)=\phi(r,t) \exp[-i m c^2 t/\hbar]$ into the Klein-Gordon equation, where $m$ is the rest mass, the non-relativistic limit is obtained by assuming that the rest mass energy $mc^2$ greatly exceeds the kinetic energy, i.e.\ $\vert i \hbar \partial \phi /\partial t \vert \ll mc^2 \phi$ (see p7 of Reference \cite{Greiner}). Thus, second order time derivatives of $\phi$ can be neglected and this leads directly to Schr\"{o}dinger's equation for a single massive particle as an approximation to the Klein-Gordon equation.  The non-relativistic ansatz should be compared with that introduced in Eq.\ (\ref{eq:DFOEansatz}) which reduces the second order Maxwell wave equation encapsulated in the DSOE to the first order Schr\"{o}dinger-like DFOE. In both cases the exponential accounts for the dominant time-dependence: this arises from the rest mass energy in the Klein-Gordon case, and in the Maxwell case from the quantities $ \sqrt{(\Gamma(t)\pm\omega_{\mathrm{av}})^{2}+\Delta^{2}}$ given in Eq.\ (\ref{eq:betadef}), i.e.\  the diagonal terms of the DSOE given in Eq.\ (\ref{eq:DSOEmatrix}). Also, second order time derivatives are likewise ignored in order to obtain the DFOE. Just as Schr\"{o}dinger's equation knows nothing about antiparticles and, indeed, conserves particle number, the Schr\"{o}dinger-like DFOE knows nothing about parametric amplification of the Maxwell field.

\acknowledgements{We thank the Natural Sciences and Engineering Research Council of Canada (NSERC) for funding.}

\begin{appendix}
  \section{Electric field in a moving dielectric }
  \label{app:NonrelativisticApproximation}
As predicted by Fresnel in 1818 \cite{Fresnel} and observed by Fizeau in 1851 \cite{Fizeau}, the apparent refractive index of a medium depends upon its velocity. This effect is in principle present in the moving membrane studied in this paper, and we shall therefore make a rough estimate of the size of the effect. 
Inside a stationary dielectric with a uniform refractive index $n_{r}$ the electric field obeys the wave equation
\begin{equation}
  \frac{\di ^2 E} {\di x^2}-\frac{n_{r}^2}{c^{2}}\frac{\di^{2}{E}}{\di t^{2}}=0.
\end{equation}
Now consider a dielectric moving with velocity $\mathbf{v}$ in the laboratory. In order to find the transformed wave equation we follow \cite{PiwnickiLeonhardt} and first rewrite the above wave equation as
\begin{equation}
  \frac{\di ^2 E} {\di x^2}-\frac{1}{c^{2}}\frac{\di^{2}{E}}{\di t^{2}}-\frac{n_{r}^2-1}{c^{2}}\frac{\di^{2}{E}}{\di t^{2}}=0.
\end{equation}
The first two terms form an invariant combination under Lorentz transformation. However, the third term is not invariant and to first order in  $\vert \mathbf{v} \vert/c$ the time derivative transforms as $\di /\di t \rightarrow  \di /\di t + \mathbf{v} \cdot \nabla $. Therefore, to this order of approximation, the electric field in the dielectric satisfies 
\begin{equation}
  \frac{\di ^2 E} {\di x^2}-\frac{n_{r}^2}{c^{2}}\frac{\di^{2}{E}}{\di t^{2}}-2 \frac{n_{r}^{2}-1}{c^{2}} \mathbf{v} \cdot \nabla \frac{\di{E}}{\di t}=0.
  \label{eq:RelMaxEq}
\end{equation}
when viewed from the laboratory frame.

The highest membrane velocity considered in this paper is $20,000$ ms$^{-1}$, and the highest membrane reflectivity is $98 \%$ for a wavenumber $k=8 \times 10^6$ m$^{-1}$. Using Eq.\ (\ref{eq:reflectivity}) for the reflectivity, we find that this implies that the $\delta$-membrane dielectric coefficient takes the value $\alpha=1.7 \times 10^{-6}$ m.
 Assuming a membrane of width $w=50$ nm, we can use the relation $\alpha = 2 w n_{r}^2$ derived in Appendix B in reference \cite{NickPaper} between $\alpha$ and the refractive index to obtain $n_{r} \approx 4$.
Armed with the refractive index, and assuming $E(x,t)=E_{0} \exp{\left [i(kx-\omega t)\right ]}$,  we can compare the order of magnitude of each term in the transformed wave equation Eq.\ (\ref{eq:RelMaxEq}). We have $\frac{\di ^2 E} {\di x^2} \sim k^2$ ;
$\frac{n^2}{c^{2}}\frac{\di^{2}{E}}{\di t^{2}} \sim n^2 k^2=16k^2$ ;  $v \frac{n^{2}-1}{c^{2}}\frac{\di}{\di x}\frac{\di{E}}{\di t} \sim \frac{v}{c}(n^2-1)k^2=0.001k^2$. We conclude that for the velocities considered in this paper the motion of the membrane only introduces a modification three orders of magnitude smaller than the standard static membrane effect  and will therefore be neglected.

\section{Initial conditions for the electric field in the adiabatic basis.}
\label{app:initialcondition}

In this appendix we find an expression for $\dot{c}_{m}(t_{0})$, where $c_{m}(t)$ is the $m^{\mathrm{th}}$ expansion coefficient of the electric field in the adiabatic basis [Eq.\ (\ref{eq:globalmodeansatz})] that is quoted at the end of Section \ref{sec:adiabaticbasisEOM}.  Our approach is adapted from that given in Appendix F.2 in Reference \cite{liningtonthesis}. We start from the two Maxwell equations $\nabla \times \mathbf{E}= -\di \mathbf{B}/ \di t$ and 
$\nabla \times \mathbf{H} = \di \mathbf{D} / \di t$ and put $\mathbf{B}(\mathbf{r},t)=\mu_{0} \mathbf{H}(\mathbf{r},t)$ and $\mathbf{D}(\mathbf{r},t)=\epsilon(\mathbf{r},t) \mathbf{E}(\mathbf{r},t)$, where $\epsilon(\mathbf{r},t)$ is the time and space dependent dielectric function appropriate to the double cavity [$n_{r}(\mathbf{r},t)= c \sqrt{\epsilon(\mathbf{r},t) \mu_{0} }$ is the refractive index]. Under the physically reasonable assumption that the time evolution of the dielectric function is much smaller than the optical frequency that determines the time evolution of the electric field, the second Maxwell equation becomes $\nabla \times \mathbf{B} \approx \epsilon \mu_{0} \di \mathbf{E} / \di t$.  In our one-dimensional system the two Maxwell equations take the forms $\di E / \di x = \di B / \di t$ and $\di B / \di x = \epsilon(x,t) \mu_{0} \di E / \di t$, respectively. The key assumption we now make is that for $t<t_{0}$ the membrane is stationary $\epsilon(x,t) \rightarrow \epsilon(x)$. This means that the adiabatic mode functions and frequencies for $t<t_{0}$ are time independent.  Next we expand the electric and magnetic field amplitudes over the adiabatic basis as 
\begin{eqnarray}
E(x,t<t_{0}) & = & \sum_{n} c_{n} U_n(x) e^{ -i\omega_n t} \\
B(x,t<t_{0}) & = & \frac{i}{c}  \sum_{n} c_{n} V_n(x) e^{ -i \omega_n t}
\label{eq:EandB}
\end{eqnarray}
where we note that the expansion coefficients are the same for both fields and that $\omega_{n}=c k_{n}$. We have also introduced $V_{n}(x)$ as the adiabatic mode functions for the magnetic field. Due to the fact that the membrane is assumed to be stationary, the adiabatic modes are not merely instantaneous eigenmodes like in the moving membrane case but are true normal modes of the double cavity that are independent of one another. This implies that the Maxwell equations must be satisfied for each mode individually and allows us to determine the relationship between the $U_{n}$ and $V_{n}$ mode functions as
\begin{eqnarray}
\frac{\di U_{n}(x)}{\di x }  & = & k_{n} V_{n}(x) \label{eq:dUdx} \\
\frac{\di V_{n}(x)}{\di x } & = &  - n_{r}^2(x) \, k_{n} U_{n}(x) . \label{eq:dVdx}
\end{eqnarray}
The second of these equations can be used to express the gradient of the total magnetic field in terms of the electric field mode functions $U_{n}$
\begin{equation}
\frac{\di B}{\di x} = - \frac{i}{c} n_{r}^2(x) \sum_{n} c_{n} k_{n} U_{n} (x) e^{ -i \omega_n t} .
\end{equation}
We now consider times infinitesimally greater than $t_{0}$ when the membrane starts moving. Inserting the above result for $\di B/\di x$ into $\di B / \di x = \epsilon(x,t) \mu_{0} \di E / \di t$ and introducing the time-dependence of all quantities gives 
\begin{eqnarray}
- i & \underset{n}{\sum} & c_{n}(t) \omega_{n}(t) U_{n} (x,t) e^{  -i\int^t_{t_0}\omega_n(t^{\prime})\mathrm{dt^{\prime}} } \\ \nonumber  & = & \frac{\di }{\di t} \left\{ \sum_{n} c_{n}(t) U_{n}(x,t) e^{ -i\int^t_{t_0}\omega_n(t^{\prime})\mathrm{dt^{\prime}}  } \right\}
\end{eqnarray}
which simplifies to
\begin{equation}
\sum_{n} \frac{\di}{\di t} \bigg\{ c_{n}(t) U_{n}(x,t) \bigg\} e^{  -i\int^t_{t_0}\omega_n(t^{\prime})\mathrm{dt^{\prime}} } = 0 .
\label{eq:initialcondition1}
\end{equation}
We emphasize that this result is only valid for $t \approx t_{0}$ since in order to derive it we assumed the results given in Eqns. (\ref{eq:dUdx}) and (\ref{eq:dVdx}) which rely on the time independence of the normal modes.

Projecting out the $m^{\mathrm{th}}$ coefficient using the orthonormality of the mode functions, we can express the relation given in Eq.\ (\ref{eq:initialcondition1})  at the initial time $t=t_{0}$ as
\begin{equation}
\dot{c}_{m}(t_{0})=- \sum_{n} P_{mn}(t_{0}) c_{n}(t_{0})
\end{equation}
where the function $P_{mn}(t)$ is defined in Eq.\ (\ref{eq:Pdefinition}). This fixes $\dot{c}_{m}(t_{0})$ for any particular choice of the initial coefficients $c_{n}(t_{0})$. 

\section{Derivation of the quantum equations of motion}
\label{app:quantumEOM}
In this appendix we give the derivation of Eqns. (\ref{eq:dQdt}) and (\ref{eq:dPdt}) which are the equations of the motion for the ``position'' $\hat{Q}_{n}$ and ``momentum'' $\hat{P}_{n}$ operators for the field modes that appear in Section \ref{sec:DCE}.  The derivation begins by taking the time derivatives of Eqns. (\ref{eq:Qn}) and (\ref{eq:Pn}) for $\hat{Q}_{n}$ and $\hat{P}_{n}$, respectively.  Taking the $\hat{Q}_{n}$ case first we have
\begin{eqnarray}
 \frac{d \hat{Q}_{n}}{dt}   & =  & \frac{1}{\sqrt{\epsilon_{0}}}    \int_{-L_{1}}^{L_{2}} dx \bigg[\frac{\partial \epsilon(x,t)}{\partial t} \hat{A}(x,t) U_{n}(x,t)  \\
   && +   \epsilon(x,t) \frac{\partial \hat{A}(x,t)}{\partial t} U_{n}(x,t) 
 + \epsilon(x,t)  \hat{A}(x,t) \frac{\partial U_{n}(x,t)}{\partial t} \bigg]  \nonumber \\
 & = & \frac{1}{\sqrt{\epsilon_{0}}}    \int_{-L_{1}}^{L_{2}} dx \bigg[ \frac{\partial \epsilon(x,t)}{\partial t} \hat{A}(x,t) U_{n}(x,t)  \nonumber \\
 &&+ \hat{\pi}(x,t) U_{n}(x,t) + \epsilon(x,t) \hat{A}(x,t) \frac{\partial U_{n}(x,t)}{\partial t} \bigg] \\
 &=& \hat{P}_{n}(t) - \sum_{m} G_{nm}(t) \hat{Q}_{m}(t)
\end{eqnarray}
which is the result given in the main text.
In going from the first equality to the second we used the definition $\pi(x,t) \equiv \epsilon(x,t) \partial A(x,t)/ \partial t$ which in turn gives $ \hat{P}_{n}(t)$ on the last line when we use the expression given in Eq.\ (\ref{eq:Pn}) for  $ \hat{P}_{n}(t)$. We also replaced $\hat{A}(x,t)$ in the other two terms by its expansion over $\hat{Q}_{m}(t)U_{m}(x,t)$ given in Eq.\ (\ref{eq:Aexpansion}) :
\begin{eqnarray}
  &&  \int_{-L_{1}}^{L_{2}}  \frac{dx}{\sqrt{\epsilon_{0}}}  \bigg[\frac{\partial \epsilon(x,t)}{\partial t} \hat{A}(x,t) U_{n}(x,t) 
  + \epsilon(x,t)  \hat{A}(x,t) \frac{\partial U_{n}(x,t)}{\partial t} \bigg] \nonumber \\
&&  = \sum_{m} \hat{Q}_{m}(t)  \int_{-L_{1}}^{L_{2}} dx \bigg[\frac{\partial}{\partial t} \frac{\epsilon(x,t)}{\epsilon_{0}} U_{m}(x,t) U_{n}(x,t) \nonumber  \\
&& \quad +  \frac{\epsilon(x,t)}{\epsilon_{0}} U_{m}(x,t) \frac{ \partial U_{n}(x,t)}{\partial t} \bigg] \\
&& = - \sum_{m}\hat{Q}_{m}(t) \int_{-L_{1}}^{L_{2}} dx \frac{\epsilon(x,t)}{\epsilon_{0}} \frac{\partial U_{m}(x,t)}{\partial t} U_{n}(x,t)  \\
&& = - \sum_{m} G_{nm} (t)\hat{Q}_{m}(t) 
\end{eqnarray}
where $G_{nm}(t)=\dot{q} g_{nm}(t)$ and $g_{nm}(t)$ is defined in Eq.\ (\ref{eq:gdefinition}). In going from the first equality to the second equality in this expression we made use of a relation obtained by differentiating the orthonormalization condition Eq.\ (\ref{eq:orthonormal}) with respect to time :
\begin{equation}
\frac{\partial}{\partial t}  \int_{-L_{1}}^{L_{2}} dx \frac{\epsilon(x,t)}{\epsilon_{0}} U_{m}(x,t)U_n(x,t) = 0.
\end{equation}

The equation of motion for $\hat{P}_{n}(t)$ is obtained similarly; differentiating Eq.\ (\ref{eq:Pn}) with respect to time yields
\begin{equation}
\frac{d \hat{P}_{n}}{dt} = \int_{-L_{1}}^{L_{2}}  \frac{dx}{\sqrt{\epsilon_{0}}} \bigg[ \frac{\partial \hat{\pi}(x,t)}{\partial t} U_{n}(x,t)+ \hat{\pi}(x,t) \frac{\partial U_{n}(x,t)}{\partial t} \bigg]. \label{eq:dPdtappendix}
\end{equation}
The first term can be reexpressed in terms of $\hat{A}(x,t)$ by using the wave equation (\ref{eq:waveeqnforA}) to write
\begin{equation}
 \frac{\partial \hat{\pi}(x,t)}{\partial t} =   \frac{1}{\mu_{0}}  \frac{\partial^2 \hat{A}(x,t)}{\partial x^2} 
\end{equation}
and replacing $\hat{A}(x,t)$ by its expansion over $\hat{Q}_{m}(t)U_{m}(x,t)$ as given in Eq.\ (\ref{eq:Aexpansion}) gives
\begin{eqnarray}
 \int_{-L_{1}}^{L_{2}}  && \frac{dx}{\sqrt{\epsilon_{0}}}   \frac{\partial \hat{\pi}(x,t)}{\partial t} U_{n} (x,t)  \nonumber \\
&& =  \sum_{m} \hat{Q}_{m}(t)  \int_{-L_{1}}^{L_{2}}  \frac{dx}{\mu_{0} \epsilon_{0}}  \frac{\partial^2 U_{m}(x,t)}{\partial x^2} U_{n}(x,t) \nonumber  \\
&& =  - \sum_{m} \hat{Q}_{m}(t) \omega_{m}^{2}(t). \label{eq:1stterm}
\end{eqnarray}
In the last step we used the time-independent wave equation (\ref{eq:Utimeindependent}) satisfied instantaneously by the adiabatic mode functions $U_{m}(x,t)$ to remove the second spatial derivative, leaving an integral corresponding to the  orthonormality  condition Eq.\ (\ref{eq:orthonormal}). The second term in Eq.\ (\ref{eq:dPdtappendix}) is treated by substituting the expansion of $\hat{\pi}(x,t)$ over $\hat{P}_{m} U_{m}(x,t)$ as given in Eq.\ (\ref{eq:piexpansion}) to give 
\begin{eqnarray}
 \int_{-L_{1}}^{L_{2}} && \frac{dx}{\sqrt{\epsilon_{0}}} \hat{\pi}(x,t) \frac{\partial U_{n}(x,t)}{\partial t} \nonumber \\
&& = \sum_{m} \hat{P}_{m}(t)  \int_{-L_{1}}^{L_{2}} dx \frac{\epsilon(x,t)}{\epsilon_{0}} U_{m}(x,t)  \frac{\partial U_{n}(x,t)}{\partial t} \nonumber \\
&& = \sum_{m}  \hat{P}_{m}(t) G_{mn}(t).  \label{eq:2ndterm}
\end{eqnarray}
The sum of Eqns. (\ref{eq:1stterm}) and (\ref{eq:2ndterm}) give the expression for $d \hat{P}_{n}/dt $ quoted in Eq.\ (\ref{eq:dPdt}) in the main part of the paper.

\section{Analytic expressions and orders of magnitude for coefficients in the quantum Hamiltonian}
\label{app:analyticexpressions}
In this appendix we outline the calculation of the coefficients $(1/\omega_{1}) d \omega_{1}/dq$,  $(1/\omega_{2}) d \omega_{2}/dq$, $\omega_{1}/\omega_{2}$, $g_{11}$, $g_{22}$, $g_{12}$, and $g_{21}$,  that appear in the two-mode quantum Hamiltonians given in Eqns.\ (\ref{eq:Lambda3}) and (\ref{eq:Lambda4}). 

We first consider $(1/\omega_{1}) d \omega_{1}/dq$, where $\omega_{1}= \omega_{\mathrm{av}}-\sqrt{\Delta^2+\Gamma^2(q)}$. Noting that $\Gamma=2 \sqrt{\gamma} q \approx 2 (\omega_{\mathrm{av}}/L) q$ the derivative can be taken. When dividing by $\omega_{1}$ we make the assumption that  $ \omega_{\mathrm{av}} \gg \sqrt{\Delta^2+\Gamma^2(q)}$ (recall that $ \omega_{\mathrm{av}}$ is assumed to be an optical frequency $\approx 2 \pi \times 10^{15}$ Hz, whereas the gap $\Delta$ at an avoided crossing, which gives the order of magnitude for $\sqrt{\Delta^2+\Gamma^2(q)}$,  is assumed to be tiny in comparison; in experiments $\Delta$ ranges from $2 \pi \times 1 $ GHz \cite{thompson08} to $2 \pi \times 0.1 $ MHz \cite{Sankey2010}.) Thus we have that 
\begin{eqnarray}
\frac{1}{\omega_{1}} \frac{d \omega_{1}}{dq} & \approx & -\frac{4 \gamma q}{\sqrt{\Delta^2+4 \gamma^2 q^2}} \times \frac{1}{\omega_{\mathrm{av}}} \nonumber \\ & \approx & -\frac{4 \omega_{\mathrm{av}} q/L^2}{\sqrt{\Delta^2+4 \omega^2 q^2/L^2}}
\end{eqnarray}
where to obtain the second line we put $\gamma \approx \omega_{\mathrm{av}}^2/L^2 $, see Eq.\ (\ref{eq:gammadef}). Within the same set of approximations, $(1/\omega_{2}) d \omega_{2}/dq$ takes exactly the same magnitude but is of opposite sign. This makes intuitive sense because after an avoided crossing one mode bends down ($\omega_{1}$) and the other bends up 
($\omega_{2}$). We can thus replace all instances of the one coefficient by the (negative) of the other.

Let us also estimate the magnitude of $(1/\omega_{1}) d \omega_{1}/dq$. In the vicinity of an avoided crossing we can replace $\sqrt{\Delta^2+\Gamma^2(q)}$ by $\Delta$ and thus
\begin{equation}
\frac{1}{2 \omega_{1}} \frac{d \omega_{1}}{dq} \sim \mathcal{O} \left( -\frac{2}{L} \frac{\omega_{\mathrm{av}}}{\Delta} \frac{q}{L} \right)
\end{equation}
which varies linearly with the membrane displacement $\Delta L = 2q$.
In the experiment by Thompson \textit{et al} \cite{thompson08}, the total length of the double cavity was $L=6.7$ cm, $\Delta=2 \pi \times 1 $ GHz, and $\omega_{\mathrm{av}}\approx \omega_{\mathrm{laser}}= 10^{15}$ rad/s. Inputting these numbers we find $(1/\omega_{1}) d \omega_{1}/dq \sim 2 \times 10^6 \times (q/L) $ m$^{-1}$. The distance the membrane needs to travel to go between two avoided crossings is $(q/L) \approx c \pi/(2 L \omega_{\mathrm{av}}) \approx 7 \times 10^{-6} $ and so this sets an upper limit on the magnitude of $(q/L)$ we are interested in. Thus, as the membrane travels from one avoided crossing to halfway to the next one $(1/\omega_{1}) d \omega_{1}/dq$ varies in magnitude from 0 to 10 m$^{-1}$. This number depends on $1/L^2$ and so in smaller cavities it would grow accordingly.

 The basic approximation underlying our calculation of $g_{ij} \equiv(1/\epsilon_{0})\int_{-L_{1}}^{L_2} dx \epsilon(x,q) U_{i}(x,q)\partial U_{j}(x,q)/\partial q$, is to assume that we can expand the adiabatic modes in terms of mode functions which are perfectly localized on the left or right side of the membrane:
 \begin{eqnarray}
 \phi_{L}^{(0)} & = & \sqrt{\frac{2}{L_{1}}} \sin \left[ n \pi \left(x/L_{1}+1 \right) \right],  \  -L_{1} \le x \le 0 \\
  \phi_{R}^{(0)} & = & \sqrt{\frac{2}{L_{2}}} \sin \left[ n \pi \left(x/L_{2}+1 \right) \right],  \  0 \le x \le L_{2}.
 \end{eqnarray}
 These modes in general differ from the diabatic modes which only equal these expressions in the limit $\Delta \rightarrow 0$. Nevertheless, as $\Delta$ is decreased one finds that these rapidly become excellent approximations for the diabatic modes, the corrections being exponentially small. Expanding the adiabatic modes as
 \begin{eqnarray}
 U_{1}= \sin \theta \ \phi_{R}^{(0)} + \cos \theta \ \phi_{L}^{(0)} \\
 U_{2}= \cos \theta \ \phi_{R}^{(0)} - \sin \theta \ \phi_{L}^{(0)} 
 \end{eqnarray}
 where $\sin \theta$ and $\cos \theta$ are given, as usual, by Eqns. (\ref{eq:sindefn}) and (\ref{eq:cosdefn}), we can obtain analytic results for $g_{11}$, $g_{22}$, $g_{12}$, and $g_{21}$. One finds that
 \begin{equation}
 g_{11}=g_{22}= \cos \theta \frac{d}{dq} \cos \theta +  \sin \theta \frac{d}{dq} \sin \theta =0
 \end{equation}
 and
 \begin{eqnarray}
 g_{12}=-g_{21} &=& \sin \theta \frac{d}{dq} \cos \theta - \cos \theta \frac{d}{dq}  \sin \theta \nonumber \\
 && = - \frac{d\Gamma(q)}{dq} \frac{\Gamma(q)\Delta}{2(\Delta^2+\Gamma^{2}(q))^{3/2}} \nonumber \\
 && \approx - \frac{\omega_{\mathrm{av}}}{L} \frac{\Gamma(q)\Delta}{(\Delta^2+\Gamma^{2}(q))^{3/2}} .
 \end{eqnarray}
 To obtain an order of magnitude estimate for $g_{12}$ we make the same assumptions as for $(1/\omega_{1}) d \omega_{1}/dq$ above and find 
 \begin{equation}
 g_{12} \sim \mathcal{O} \left( -\frac{2}{L} \left(\frac{\omega_{\mathrm{av}}}{\Delta}\right)^2 \frac{q}{L} \right)
 \end{equation}
 which is a factor of $\omega_{\mathrm{av}}/\Delta \approx 10^5 $ \emph{bigger} than $(1/\omega_{1}) d \omega_{1}/dq$.
 
 Finally, we need the factors $\sqrt{\omega_{1}/\omega_{2}}$ and $\sqrt{\omega_{2}/\omega_{1}}$ which multiply $g_{12}$ and $g_{21}$, respectively, in the main Hamiltonian given in Eqns.\ (\ref{eq:QuantumHamiltonian2}) and (\ref{eq:Lambda1}). We have
 \begin{eqnarray}
 \sqrt{\frac{\omega_{2}}{\omega_{1}}} & = &\sqrt{\frac{\omega_{\mathrm{av}}+\sqrt{\Delta^2+\Gamma^2}}{\omega_{\mathrm{av}}-\sqrt{\Delta^2+\Gamma^2}}} \nonumber \\
 & = & 1 + \frac{\sqrt{\Delta^2+\Gamma^2}}{\omega_{\mathrm{av}}}+\frac{1}{2}\left(  \frac{\sqrt{\Delta^2+\Gamma^2}}{\omega_{\mathrm{av}}} \right)^2 + \cdots
 \end{eqnarray}
 and
 \begin{eqnarray}
 \sqrt{\frac{\omega_{1}}{\omega_{2}}} & = &\sqrt{\frac{\omega_{\mathrm{av}}-\sqrt{\Delta^2+\Gamma^2}}{\omega_{\mathrm{av}}+\sqrt{\Delta^2+\Gamma^2}}} \nonumber \\
 & = & 1 - \frac{\sqrt{\Delta^2+\Gamma^2}}{\omega_{\mathrm{av}}}+\frac{1}{2}\left(  \frac{\sqrt{\Delta^2+\Gamma^2}}{\omega_{\mathrm{av}}} \right)^2 + \cdots
 \end{eqnarray}
 The corrections to unity, in powers of $\sqrt{\Delta^2+\Gamma^2}/\omega_{\mathrm{av}}$,   are small. However, the first correction must be retained to be consistent with other terms involving$(1/\omega_{1}) d \omega_{1}/dq$ which is a factor  $\Delta/\omega_{\mathrm{av}}$ smaller than $g_{12}$ and $g_{21}$.

\end{appendix}

\end{document}